\documentclass[twocolumn]{revtex4}
\pdfoutput=1
\usepackage{amsmath}
\usepackage{amsfonts}
\usepackage{hyperref}
\usepackage{amssymb}
\usepackage{graphicx}
\usepackage{natbib}
\usepackage{bm}
\usepackage{outlines}
\usepackage{color}
\usepackage{verbatim}

\newcommand{\om}{\omega}

\newcommand{\Om}{\Omega}

\newcommand{\p}{\partial}

\newcommand{\avg}[1]{\left\langle #1 \right\rangle}

\newcommand{\sech}{\operatorname{sech}}
\newcommand{\cn}{\operatorname{cn}}
\newcommand{\sn}{\operatorname{sn}}
\newcommand{\dn}{\operatorname{dn}}
\begin{document}
\title{ 
Phase-space mixing in dynamically unstable, 
integrable few-mode quantum systems
}
\author{R.~Mathew}
\affiliation{Joint Quantum Institute, 
University of Maryland and 
National Institute of Standards and Technology,
College Park, Maryland 20742, USA}
\author{E.~Tiesinga}
\affiliation{Joint Quantum Institute and Joint Center for Quantum Information and Computer Science,
National Institute of Standards and Technology and University of Maryland, Gaithersburg, Maryland 20899, USA}

 \begin{abstract}

Quenches in isolated quantum systems are currently a subject of
intense study. Here, we consider quantum few-mode systems that
are integrable in their classical mean-field limit and become dynamically
unstable after a quench of a system parameter.  Specifically, we study
a Bose-Einstein condensate (BEC) in a double-well potential
and an antiferromagnetic spinor BEC constrained to a single spatial mode.
We study the time dynamics after the quench within the truncated Wigner
approximation (TWA) and find that system relaxes to a steady state due to
phase-space mixing. Using the action-angle formalism and a pendulum as
an illustration, we derive general analytical expressions for the time
evolution of expectation values of observables and their long-time limits.
We find that the deviation of the long-time expectation value from its
classical value scales as $1/O(\ln N )$, where $N$ is the number
of atoms in the condensate.  Furthermore, the
relaxation of an observable to its steady state value is a damped
oscillation and the damping is Gaussian in time.
We confirm our results with numerical TWA simulations.

\end{abstract}
\maketitle
\section{Introduction}

The advent of precise experimental control in ultracold atomic
systems has motivated theoretical study in non-equilibrium dynamics
in isolated quantum systems \cite{bloch_many-body_2008}. For generic
Hamiltonian systems the expectation value of a local observable at long times after
a quench, a sudden change in a control parameter, is described by a
Gibbs ensemble \cite{dziarmaga_dynamics_2010,dalessio_quantum_2016}.
However, for integrable systems, a special class of Hamiltonian systems, 
the long-time behavior is instead believed to be described
by a generalized Gibbs ensemble \cite{dalessio_quantum_2016}. This
important role of integrability on the time dynamics has been demonstrated
experimentally \cite{kinoshita_quantum_2006,langen_experimental_2015}.
Integrable systems are of much theoretical interest as they are amenable
to exact analytic treatment. A classical integrable system can be solved
using action-angle variables \cite{arnold}, while a quantum integrable
system is solvable by the Bethe ansatz \cite{korepin1997quantum}.

A mean-field approximation can be applied to a 
bosonic system with a macroscopically-occupied mode.
The time dynamics of the
system is then governed by a classical Hamiltonian and described by
classical trajectories in its phase space.  For a weakly interacting
Bose-Einstein condensate (BEC) this classical trajectory is a solution
of the time-dependent Gross-Pitaevskii equation for the order parameter
with continuous spatial degrees of freedom \cite{pethick_2008}.  In certain cases, it
is sufficient to describe a bosonic system with just a few degrees of
freedom. Some examples are a BEC in a double-well potential \cite{smerzi_quantum_1997},
a spin-1 spinor BEC within the single-mode approximation (SMA) 
\cite{law_quantum_1998,zhang_coherent_2005}
and a few-site Bose-Hubbard model with a large occupation per site
\cite{mossmann_semiclassical_2006,itin_semiclassical_2011,satija_soliton_2013}.

A bosonic system becomes dynamically unstable when it
is prepared by a quench at a saddle point in its phase
space. 
Dynamical instabilities have been predicted for
vortices in trapped BECs
\cite{pu_coherent_1999,skryabin_instabilities_2000,kawaguchi_splitting_2004}, 
superfluid flow of BECs in optical lattices
\cite{wu_landau_2001,menotti_superfluid_2003}
and BECs in cavities \cite{moore_quantum_1999}. These predictions
have been experimentally observed 
\cite{shin_dynamical_2004,fallani_observation_2004,cristiani_instabilities_2004,
ferris_dynamical_2008,schmidt_dynamical_2014}.
The instability is also used as an experimental
route for the generation of squeezed states
\cite{esteve_squeezing_2008,leslie_amplification_2009,klempt_parametric_2010,hamley_spin-nematic_2012}.
A mean-field description is then insufficient and quantum
fluctuations need to be included. Quantum corrections can be
(partially) included by using the truncated Wigner approximation (TWA)
\cite{sinatra_truncated_2002,blakie_dynamics_2008,polkovnikov_phase_2010},
which models the dynamics of the Wigner distribution in the phase
space. The TWA has been used to numerically study the effects of thermal
fluctuations on a BEC \cite{sinatra_truncated_2002}, quenches in spinor
condensates \cite{sau_spin_2010,barnett_prethermalization_2011} and
superfluid flow \cite{mathey_decay_2014}.

In this paper we analytically study the time dynamics of two integrable
few-mode quantum systems within the truncated Wigner approximation after
a quench of a parameter that makes the systems dynamically unstable.
Our paper is set up as follows. We introduce dynamical instability in
bosonic systems in Sec.~\ref{sec:dyn} and TWA in Sec.~\ref{sec:twa}.
We define the integrability of classical Hamiltonians, which govern the
mean-field limit of these systems, and introduce action-angle coordinates
in Sec.~\ref{sec:classical}.  Section \ref{sec:mixing} introduces the
concept of mixing in phase-space due to time evolution and describes how this
mixing leads to relaxation of an observable to a steady-state value.
Using the pendulum as an illustrative example, we derive general results
for long-time expectation value of an observable in Secs.~\ref{sec:sptx}
and \ref{sec:long_time_avg} and the time dynamics of relaxation of this
expectation value in Sec.~\ref{sec:relax}.  We apply these results to
the case of a condensate in a double-well potential (the double-well
system) in Sec.~\ref{sec:DW} and a spin-1 BEC described by a
single spatial mode in Sec.~\ref{sec:spinor}. Finally,  we conclude in
Sec.~\ref{sec:conclusion}.

\section{Dynamical instability}
\label{sec:dyn}

The mean-field equations of motion of an isolated quantum bosonic system
are equivalent to Hamilton's equations of motion of a classical 
system. The mean-field ground state is a stable equilibrium phase-space
point, where the classical Hamiltonian has a minimum.  On the other
hand, a dynamically unstable state corresponds to a saddle point of this
Hamiltonian. Such an unstable state can be prepared by starting from 
a minimum of the initial Hamiltonian and then quenching
a system parameter to change this point to a saddle point of the final
Hamiltonian. As an example, consider the quantum oscillator $H_0 =
(\hat p^2+ \hat x^2)/2$, where $\hat x$ and $\hat p$ are the canonical
position and momentum operators, respectively.  Here, we have set $\hbar$
and the natural frequency of the oscillator to one.  Its mean-field ground
state is the phase-space point $(x_c, p_c)= (0, 0)$, where $x_c=\avg{\hat
x}$, $p_c = \avg{\hat p}$, and $\avg{\dots}$ is the average over a quantum
state.  We make the state dynamically unstable by suddenly changing to
the Hamiltonian $H_1 = (\hat p^2- \hat x^2)/2$.  Under the mean-field
equations of motion a dynamically unstable point is stationary.  Thus,
$x_c(t)=0$ and $p_c(t) =0$ holds for all times.  In contrast, quantum
evolution under $H_1$ leads to exponential growth in the unstable mode
\cite{pethick_2008}.  In fact, following the language of quantum optics,
$H_1 \propto \hat a\hat a + \hat a^\dagger \hat a^\dagger$ leads to
single-mode squeezing, where $\hat a(\hat a^\dagger) = (\hat x \pm i\hat
p)/\sqrt{2}$ is the annihilation (creation) operator of the mode.

\section{Truncated Wigner Approximation}
\label{sec:twa}

The time evolution of a dynamically-unstable system can be studied using
the truncated Wigner approximation (TWA)~\cite{sinatra_truncated_2002}.
It incorporates the leading order quantum corrections to the mean-field
equations of motion \cite{polkovnikov_quantum_2003}. In the TWA  a Wigner
distribution function $F(\mathbf x, \mathbf p, t)$ time evolves
under classical Hamilton's equations, in contrast to the mean-field
approximation where the evolution of a single phase-space point $(\mathbf
x(t), \mathbf p(t))$ is studied. Here, $\mathbf x =(x_1, \dots, x_n)$
and $\mathbf p =(p_1, \dots, p_n)$ are canonical position and momentum
coordinates for a classical mean-field Hamiltonian system with $n$
degrees of freedom.  The initial distribution, $F_0(\mathbf x, \mathbf
p)$, is the Wigner transform~\cite{case_wigner_2008} of the prequench
quantum ground state or any approximation thereof. 

For an observable $\mathcal O(\mathbf x, \mathbf p)$, we define its 
evolution 
$\mathcal O(t) \equiv \mathcal O(\mathbf x(t), \mathbf p(t))$ along
a trajectory $(\mathbf x(t), \mathbf p(t))$ with initial conditions
$(\mathbf x_0, \mathbf p_0)$. 
The expectation value of $\mathcal O(t)$ over all trajectories is
\begin{align}
    \avg{ {\mathcal O}(t)} &= 
    \int_\Om d\mathbf x d\mathbf p \, \mathcal O(\mathbf x, \mathbf p)
                F(\mathbf x, \mathbf p, t)\nonumber \\
    & = \int_\Om d\mathbf x_0 d\mathbf p_0 \, \mathcal O(t)
                F_0(\mathbf x_0, \mathbf p_0),
    \label{eq:def_avg} 
\end{align}
with measures $d\mathbf x= dx_1\cdots dx_n$, $d\mathbf p= dp_1\cdots
dp_n$ and the integral is over all phase space $\Om$. 
The distribution satisfies $\int_\Omega d\mathbf x d\mathbf p \, F(\mathbf
x, \mathbf p, t)=1$ for all $t$ in accordance with Liouville's
theorem \cite{arnold}. 

\section{Classical integrable systems} 
\label{sec:classical}

In classical mechanics a Hamiltonian system with $n$ degrees of freedom
is integrable if there exist $n$ mutually commuting (with respect to the
Poisson bracket) conserved quantities \cite{arnold}. Then a trajectory
in the $2n$ dimensional phase-space lies on an $n$-dimensional torus. For
an integrable system, the coordinates $(\mathbf x, \mathbf p)$ can be
transformed to canonical coordinates called actions $\mathbf I = (I_1,
\dots, I_n )$ and angles $\bm\varphi = (\varphi_1, \dots, \varphi_n)$,
such that Hamiltonian $H$ is independent of $\bm\varphi$.  
Crucially, $( \mathbf I, \bm \varphi)$ and 
$(\mathbf I, \bm\varphi + 2\pi\mathbf m)$ correspond to the same phase-space
point, where $\mathbf m = (m_1, \dots, m_n)$ is a vector of integers.
In these coordinates, the Hamilton's equations are
\begin{equation} 
    \dot I_i = -\frac{\p H(\mathbf I)}{\p \varphi_i} = 0\, , 
    \quad 
\dot \varphi_i = \frac{\p
H(\mathbf I)}{\p I_i} \equiv \omega_i(\mathbf I),
    \label{eq:Ham_eqns}
\end{equation} 
for all $i \in \{1, \dots, n\}$.
The frequencies $\omega_i(\mathbf I)$ only depend 
on $\mathbf I$. Hence, the actions are conserved quantities
and the time evolution of the angles has the simple form 
\begin{equation}
    \bm\varphi(t) = \bm\omega(\mathbf I) t + \bm\varphi_0,
    \label{eq:angle_evolution}
\end{equation} 
where $\bm \om(\mathbf I) = (\om_1(\mathbf I), \dots, \om_n(\mathbf I))$
and $\bm\varphi(0) =\bm\varphi_0 $.

For our Hamiltonian systems action-angle coordinates are not globally
defined. Instead, they are defined on disjoint regions of 
$\Omega$ by maps from each such region $R$ to
$\mathcal I_R \otimes \mathcal J$,
where $\mathcal I_R \subset \mathbb R^n$ and $\mathcal J = [0, 2\pi]^{\otimes n}$
are the spaces spanned by the actions and angles, respectively.
We then construct
distribution functions $f_R(\mathbf I, \bm\varphi,
t) = (2\pi)^n F\left(\mathbf x(\mathbf I, \bm\varphi), \mathbf p(\mathbf I,
\bm\varphi), t \right)$ for $(\mathbf x, \mathbf p)\in R$ with normalization
$\sum_R \int_{\mathcal I_R} d\mathbf I
\int_{\mathcal J} d\bm\varphi/(2\pi)^n \, f_R(\mathbf I, \bm \varphi, t)
= 1$. The latter follows from  
the fact that canonical transformations have a unit Jacobian. The 
distribution $f_R(\mathbf I, \bm \varphi, t)$ is periodic in $\bm\varphi$
and 
evolves as $f_R(\mathbf I, \bm\varphi,  t) = f_{0, R}(  \mathbf I,
\bm\varphi-\bm\omega t)$,
where $f_{0, R}(\mathbf I, \bm \varphi) = 
f_R(\mathbf I, \bm \varphi, 0)$ is the initial distribution.
Moreover, Eq.~\ref{eq:def_avg} becomes
\begin{align}
    \avg{ {\mathcal O}(t)} = \sum_R
     \int_{\mathcal I_R} d\mathbf I 
     \int_{\mathcal J} \frac{d\bm\varphi}{(2\pi)^n} \,
    f_R(  \mathbf I, \bm\varphi, t) 
    \mathcal O_R( \mathbf I, \bm\varphi)
    \label{eq:def_avg2}
    \\
    = 
    \sum_R
    \int_{\mathcal I_R} d\mathbf
    I \int_{\mathcal J} \frac{d\bm\varphi_0}{(2\pi)^n} \, 
    f_{0,R}(\mathbf I, \bm\varphi_0) 
    \mathcal O_R(\mathbf I, \bm\varphi(t)),
    \label{eq:def_avg3}
\end{align}
where $\mathcal O_R(\mathbf I, \bm\varphi)$ is the functional 
form of the observable in region $R$.

\section{Phase-space mixing}
\label{sec:mixing}

\begin{figure}
  \begin{center}
    \includegraphics[scale=0.58]{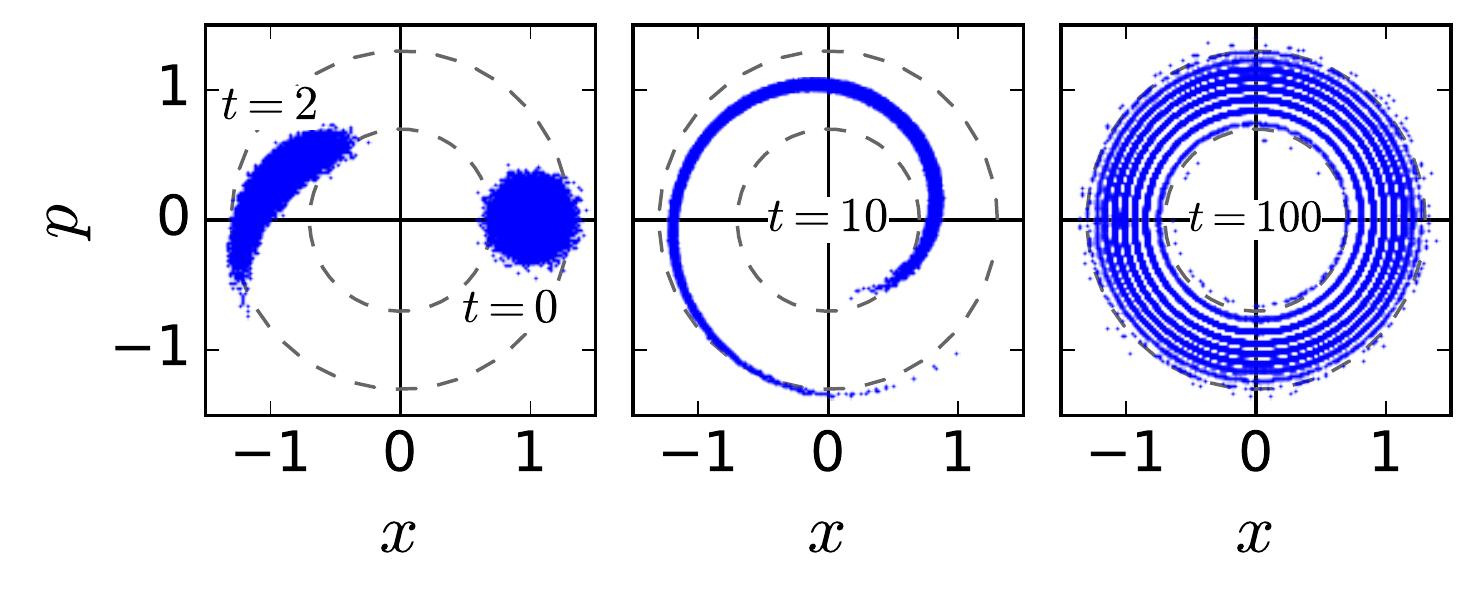}
  \end{center}
  \caption{Phase-space mixing for an anharmonic oscillator with $\varepsilon=1$. 
    Panels show the Wigner distribution  $F(x,p,t)$ in phase-space $(x, p)$
    at times $t=0$, 2, 10 and 100. Initially
    $F(x,p,t=0)$ is a 2D Gaussian with standard deviation $\sigma = 0.1$
    localized around $(x, p) = (1, 0)$. Approximately $99.7\%$ of the points 
    lie within the two dashed circles.
  }
  \label{fig:phase_mixing}
\end{figure}

A distribution function that is initially localized around a
phase-space point typically stretches, tangles and disperses over
the accessible phase space. This mixing in phase space has been
studied in plasma physics \cite{hammett_fluid_1992} and astrophysics
\cite{tremaine_geometry_1999}. We illustrate this concept using
an anharmonic oscillator.  Its Hamiltonian $H = r^2/2 +
\varepsilon r^4$ is integrable, where $r^2 = p^2 + x^2$ and we have set the mass
and the natural frequency of the oscillator to unity. In this case
the action-angle coordinates are globally defined. The action $I$ is
a function of $r$ and the angle $\varphi$ is the polar angle in the
$(x,p)$ plane. Points with different $r$ rotate around the origin at
different frequencies $\om(I)$ and the distribution function stretches as
shown Fig~\ref{fig:phase_mixing}. Eventually, the distribution spreads
uniformly and mixes in the compact coordinate $\varphi$, while remaining
localized in $r$ and $I$.

For a general integrable system, the frequencies $\bm \omega(\mathbf I)$
depend nontrivially on $\mathbf I$. Hence, the distribution will eventually
mix in $\bm \varphi$. It is important to realize that as the distribution
function mixes in phase space fine-scale structures must develop in order
to conserve the phase-space volume as required by Liouville's theorem.
For the anharmonic oscillator evolution leads to tightly wound spirals
as shown in the third panel of Fig.~\ref{fig:phase_mixing}.

Phase-space mixing simplifies the evaluation of the long-time expectation
value of an observable.  Experimentally-accessible observables are
typically smooth functions of the phase-space coordinates. Then the distribution function 
with its fine-scale structures can be coarsened, i.e.,
in Eq.~\ref{eq:def_avg2} we can replace $f_R(\mathbf I, \bm \varphi, t)$ by
the time-independent distribution \cite[\S 1]{Land10}
\begin{equation}
    \bar f_R(\mathbf I) \equiv 
    \int_{\mathcal J} 
    \frac{d\bm\varphi}{(2\pi)^n}  
    f_R(\mathbf I, \bm\varphi, t)
    =
    \int_{\mathcal J} 
    \frac{d\bm\varphi}{(2\pi)^n}  
    f_{0, R}(\mathbf I, \bm\varphi).
    \label{eq:reduced_dist} 
\end{equation} 
Consequently, the expectation value at long times becomes
\begin{equation}
    \lim_{t\to \infty} \avg{ {\mathcal O}(t)}
    = \sum_R\int_{\mathcal I} d\mathbf I\, 
    \bar f_R(\mathbf I)
  \int_{\mathcal J}\frac{d\bm\varphi}{(2\pi)^n}  
    \mathcal O_R(\mathbf I, \bm\varphi).
\label{eq:long_time_avg}
\end{equation} 
Thus, the long-time expectation value of 
an observable is given by the average over the accessible phase space
weighted by $\bar f_{R}(\mathbf I) $.

\section{Dynamics near a separatrix }\label{sec:sptx}

The description of the time evolution of the initially-localized Wigner
distribution following dynamical instability for our double-well and
spin-1 boson systems with a four- and six-dimensional phase space,
respectively, must include a study of separatrices.
As we will show in
Sec.~\ref{sec:DW} and \ref{sec:spinor} their dynamics is
controlled by a two-dimensional subspace $\Omega_{\rm 2D}$ spanned
by canonical coordinates $x_1$ and $p_1$. This subspace
contains a single saddle point that is connected to itself by one or
more trajectories, known as separatrices. In fact,
there are two separatrices and one separatrix for the double-well and
spin-1 Bose system, respectively.  The frequency $\omega_1(\mathbf
I)$ associated with a trajectory in $\Omega_{\rm 2D}$ goes to zero as its
starting point approaches the saddle point.  
In fact, near the saddle point $\omega_1$ varies sharply with $\mathbf I$,
which leads to phase-space mixing in $\Omega_{\rm 2D}$. 
The other frequencies
$\omega_i$ for $i\neq 1$ are slowly-varying near the saddle point
and the distribution along the corresponding angles remain localized
over the timescale for phase-space mixing in $\Omega_{\rm 2D}$.  
In this and the next section we discuss general features of 
trajectories and observables 
in the phase space region near a separatrix.
We develop this discussion using a simple pendulum, an integrable
system with a two-dimensional phase space 
containing a single saddle point and two separatrices \cite[\S 22.19]{dlmf}.

The Hamiltonian of a simple pendulum is 
\begin{equation}
    H_{\rm pend} = \frac{p^2}{2} + 1 - \cos\theta, 
    \label{eq:Hpend}
\end{equation}
where $p$
is the momentum and $\theta \in [-\pi, \pi]$ is the 
angular position, where $\theta=\pm \pi$ are identical (we have set the
pendulum's length and acceleration due to gravity to one).  The point $(\theta, p) =
(0, 0)$ corresponds to the stable equilibrium, while $(\theta, p) =
(\pi, 0)$ is its sole saddle point and corresponds to a stationary
upright pendulum. Around the saddle point $H_{\rm pend} \sim 2 + (p^2 - x^2)/2$, 
where $x = (\theta -\pi)\mod 2\pi$.

\begin{figure}
  \begin{center}
    \includegraphics[scale=0.5]{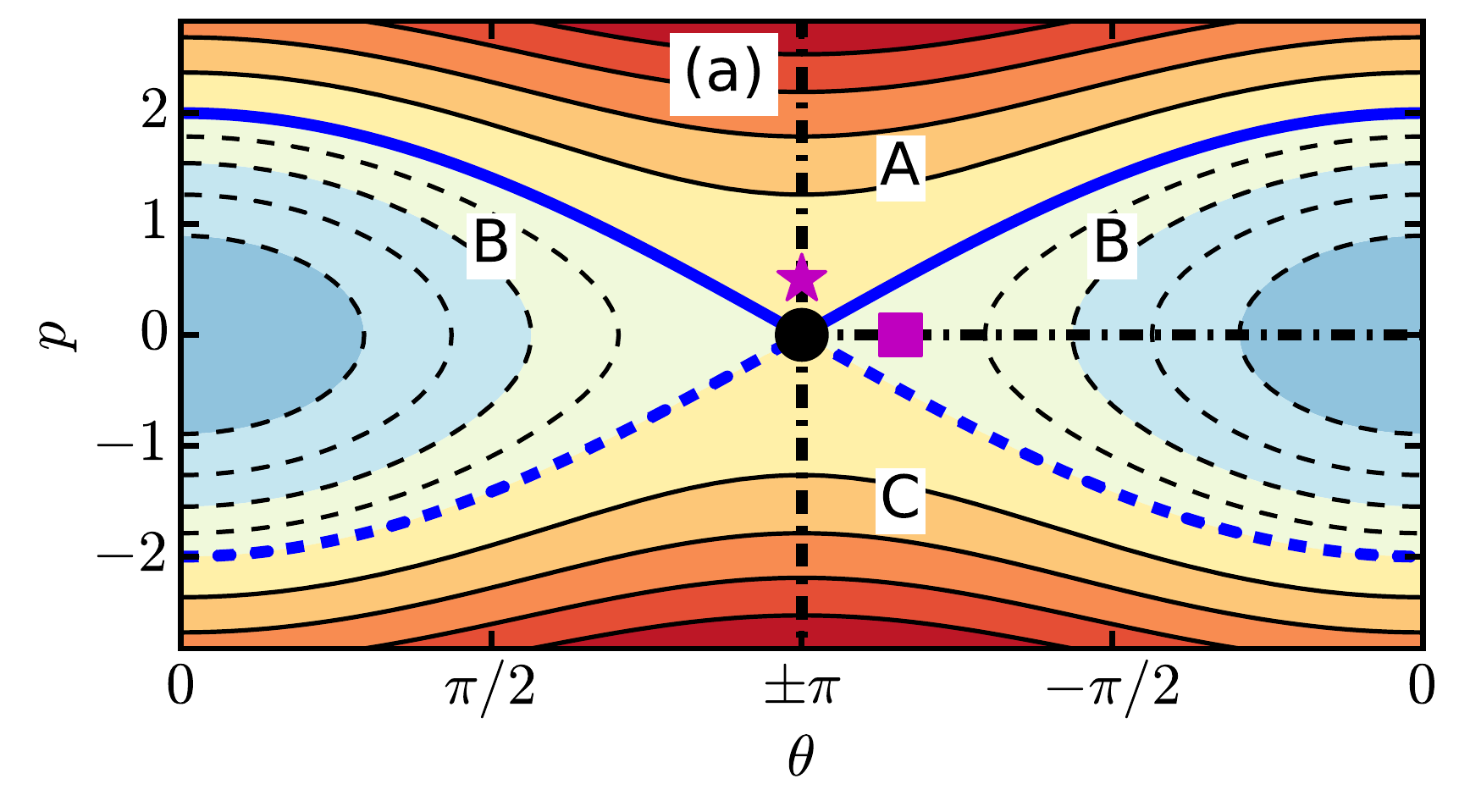}
    \includegraphics[scale=0.5]{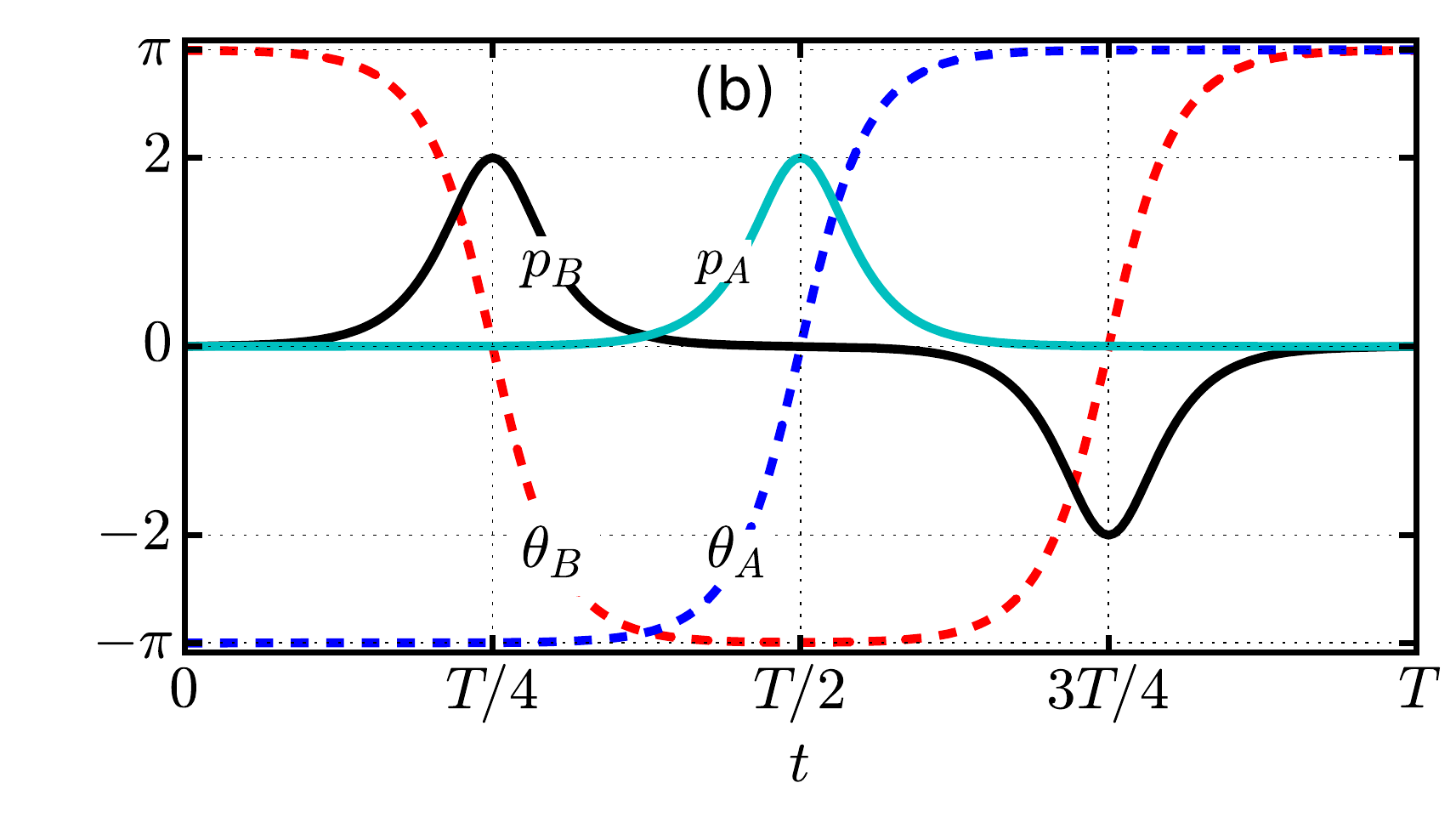}
  \end{center}
  \caption{  
  (a) Equal-energy contours in the phase space $(\theta, p)$ of a simple pendulum.
  The phase-space is a cylinder as the lines $\theta=-\pi$ and $\pi$ are equivalent. 
  The saddle point is at $(\theta, p)=(\pi, 0)$. Separatrices are 
  thick solid blue ($S+$) and thick dashed blue ($S-$) lines, and divide the phase space into 
  libration ($A$ and $C$) and rotating ($B$) regions. 
 For each region the thick dashed-dotted black line defines action-angle coordinate $\varphi_1 = 0$.
  (b) Trajectories starting near the saddle point as a function of 
  time $t$ for a single period $T$. 
  Both a rotational $(\theta_A(t), p_A(t))$ and a librational $(\theta_B(t), p_B(t))$
  trajectory are shown.
  The rotational trajectory lies in region $A$ and starts from phase-space point 
  denoted by a star in panel (a). The librational trajectory lies in region $B$ and 
  starts from the square in panel (a).
  } 
  \label{fig:pendulum_phase_space}
\end{figure}

Figure \ref{fig:pendulum_phase_space}(a) shows the equal-energy contours
in the phase space of the pendulum.  Two separatrices, $S+$ and $S-$,
divide the phase space into three regions, denoted by $A$, $B$ and $C$,
with two distinct kinds of periodic motions: libration and rotation.
Libration, confined to region $B$, is an oscillation where $\theta$ is
bounded and does not pass the inverted position, $\theta=\pi$.  Its time
period is $T_{\rm lib} = 4 K(k)$, where $K(k)$ is the elliptic integral
of the first kind \cite{dlmf}, the modulus $k= \sqrt{\mathcal E/2}$ and $\mathcal E$
is the energy.  Rotation is an unbounded motion in regions $A$ or $C$,
where the pendulum passes the inverted position.  Its time period is
$T_{\rm rot} = 2 k K(k)$, where $k = \sqrt{2/\mathcal E}$. Explicit expressions
of libration and rotation motion are given in App.~\ref{app:pendulum}.

On the separatrices the period is infinite and, hence, the action
angle coordinates $(I_1, \varphi_1)$ are not defined.  Thus, 
a saddle point precludes the existence of global
action-angle coordinates. They are, however,
defined separately in each of the three regions. Although,
the explicit form of $I_1$ and $\varphi_1$ in terms of $p$ and $\theta$
is known \cite{brizard_action-angle_2011}, it is not required for our
analysis. We will need the location where $\varphi_1$ is zero
along an equal-energy contour. We define it to be a point near the saddle
point where $|p|$ is minimal. This condition is unique for regions $A$ and
$C$. In region $B$ there are two such points and we choose
the point where $\theta >0$. As the travel time between the two points
is a half the period, $\varphi_1 = \pi$ for the other point. Our choice
of $\varphi_1 = 0$ is shown in Fig.~\ref{fig:pendulum_phase_space}(a)
as dashed-dotted lines originating from the saddle point.

We remark on the properties of solutions on the separatrix, which will
be useful later.  The two solutions that vary significantly only around $t=0$
and for which $\theta(t=0)=0$ are given by
\begin{equation}
    \theta_{\mathrm S \pm}(t)=\pm 2\arcsin(\tanh t), \quad
    p_{\mathrm S \pm}(t) = \pm 2\sech(t).
    \label{eq:pend_sptx}
\end{equation}
Note that $p_{S\pm}(t)$ is well approximated by a bump function 
(also known as a test function \cite{Gelfand})
that is nonzero in a finite domain, called the support, and vanishes
outside its support. Moreover, an observable $\mathcal O(t)$  on the
separatrix is (well approximated) by a constant plus a bump function,
as long as it is smooth in both $p$ and $\theta$ and periodic in $\theta$.

Trajectories $(\theta(t), p(t))$ that start near one of the separatrices
spend most of their time (within a period) near the saddle point as
shown with two examples in Fig.~\ref{fig:pendulum_phase_space}(b).
Changes in $\theta(t)$ and $p(t)$ from their saddle-point value 
are to good approximation equal to corresponding changes
along trajectories on one or more of the separatrices.  For example,
for the rotation trajectory in Fig.~\ref{fig:pendulum_phase_space}(b)
the momentum is $p_A(t) = p_{\mathrm S+}(t-T_{\rm rot}/2) $ for
$t \in [0, T_{\rm rot})$, while  for the libration trajectory in
Fig.~\ref{fig:pendulum_phase_space}(b) the momentum is $p_B(t) = p_{\mathrm
S+}(t- T_{\rm lib}/4) + p_{\mathrm S-}(t- 3 T_{\rm lib}/4)$ for $t \in
[0, T_{\rm lib})$.  In fact, the momentum along any trajectory starting
near the saddle point in region $R=A$, $B$ or $C$, respectively, can
be written as \begin{equation} p_R(t) \sim \sum_{n=-\infty}^\infty
\left[\sum_{s=\{S\pm\}} \chi_R(s)\, p_s\left(t-t_{0, R}(s) - n T_R
\right)\right],
    \label{eq:p_t}
\end{equation} where  the sum over $n$ defines the momentum for all $t$
(rather than a single period) and indicator functions $\chi_R(s)$ 
are either zero or one. For the pendulum $\chi_A(S+)$, $\chi_B(S+)$,
$\chi_B(S-)$ and $\chi_C(S-)$ are one; others are zero.  The time shift
$t_{0, R}(s)\in [0,T_R)$ and period $T_R$ are determined by the starting
point, where  $T_R =T_{\rm rot}$ and $T_{\rm lib}$ for $R = A, C$ and $R =
B$, respectively.  Thus, $p_R(t)$ is a sum over periodically occurring,
non-overlapping bump functions whose support is much smaller than the
time period.

The asymptotic symbol $\sim$ in Eq.~\ref{eq:p_t} and elsewhere in this paper
implies that either the trajectories start close to the saddle point or 
the averages are over a Wigner distribution that is initially localized around 
the saddle-point and whose initial width goes to zero.
We also reserve the word asymptotic for these two cases, unless 
otherwise stated.

\section{Long-time expectation value}
\label{sec:long_time_avg}

We now derive the long-time expectation value of observables $\langle\mathcal
O(t)\rangle$ that are smooth functions of the canonical coordinates
$(\mathbf x, \mathbf p)$ and depend only on the single action-angle
coordinate
$\varphi_1$ of the subspace $\Omega_{\rm 2D}$ in which the system
undergoes phase-space mixing. For periodic coordinates, like angle
$\theta$ of the pendulum, we restrict the observables to be periodic
in $\theta$. These constraints are not severe as many physically
interesting observables have these properties.

The first step is to write the asymptotic form of observable $\mathcal
O_R(t)$ in region $R$, along a trajectory that comes close to the saddle
point, in terms of its value along the separatrix trajectories $(x_{1,
s}(t), p_{1, s}(t))$ in subspace $\Omega_{\rm 2D}$. Here $s$ labels
separatrices.  (For the pendulum $s \in \{S+, S-\}$.)  We define $\mathcal
O_s(t) = \mathcal O(x_{1,s}(t),p_{1,s}(t))$ and realize that $\mathcal O_s(t)
= \mathcal O_{\rm sp} + \mathcal D_s(t)$, where $\mathcal O_{\rm sp}$ is
the value of the observable at the saddle point and $\mathcal D_s(t)$ is a
bump function localized around $t=0$. 
Similarly, we decompose $\mathcal
O_R(t) =\mathcal O_{\rm sp} + \mathcal D_R(t)$, where $\mathcal D_R(t)$
is a series of periodically occurring, non-overlapping bump
functions. Then, similar to Eq.~\ref{eq:p_t}, we write
\begin{equation}
    \mathcal O_R(t) \sim 
    \mathcal O_{\rm sp} + 
    \sum_{n=-\infty}^\infty \sum_s \chi_R(s) \mathcal D_s(t-t_{0,R}(s) - n T_R).
    \label{eq:Ot}
\end{equation}
The indicator functions $\chi_R(s)$ are system dependent and the sum $s$ 
is over one or more separatrices.  

To compute the long-time limit of $\avg{\mathcal O(t)}$ using
Eq.~\ref{eq:long_time_avg} we need to evaluate the integral over
angle $\varphi_1$. (Those over $\varphi_j$ for $j>1$  evaluate to
unity for allowed observables.) We transform this integral to one over time by
choosing a reference trajectory that starts near the saddle point with
$\varphi_1(0) =0$. For the pendulum, two such trajectories are shown in
Fig.~\ref{fig:pendulum_phase_space}(b). Then $\varphi_1(t) = \omega_1 t$
and 
\begin{align}
    \label{eq:O_angle_avg}
    \int_0^{2\pi} \frac{d\varphi_1}{2\pi} 
    \mathcal O_R( {\bf I}, \varphi_1)
    &\sim
    \mathcal O_{\rm sp}
    + \sum_s  \chi_R(s) \frac{\omega_1(\mathbf I)}{2\pi}\\
    & \times \sum_{n=-\infty}^\infty
    \int_{0}^{T_R}dt\, 
    \mathcal D_s(t- t_{0, R}(s) -nT_R).
     \nonumber
\end{align}
For $n=0$ the integrand $\mathcal D_s(t-t_{0, R}(s))$ is localized
around $t = t_{0, R}(s) \in (0, T_R)$.  Its support is enclosed by the
integration bounds $t=0$ and $T_R$ as the reference trajectory is near
the saddle point at these times. For $n \ne 0$ there is no overlap
between the support and the integration interval; hence,
the integral is zero.
We extend the integration limits
of $t$  to $(-\infty, \infty)$ for the surviving $n=0$ term 
and find
\begin{align}
    \int_0^{2\pi} \frac{d\varphi_1}{2\pi} 
    \mathcal O_R( {\bf I}, \varphi_1)
    &\sim
    \mathcal O_{\rm sp}
    + \sum_s  \chi_R(s) \frac{\omega_1(\mathbf I)}{2\pi}\\
    & \quad\quad\quad\quad\quad \times\int_{-\infty}^\infty  dt\,
     \mathcal D_s(t).
     \nonumber
\end{align}
Substituting this expression in Eq.~\ref{eq:long_time_avg}
the long-time average becomes
\begin{align}
\lim_{t\to \infty} \avg{ {\mathcal O}(t)} 
 \sim
 \mathcal O_{\rm sp} + \sum_R\frac{\avg{\omega_1}_R }{2\pi}
        \left[
        \sum_s \chi_R(s)\int_{-\infty}^{\infty}dt\, \mathcal D_s(t)
        \right],
        \label{eq:long_time_avg_final}
\end{align}
where the average  frequency $\avg{\omega_1}_R=\int_R d\mathbf I\,
\bar f_R(\mathbf I)\,\omega_1(\mathbf I)$ and the expression in
the square bracket is independent of the distribution.  Equation
\ref{eq:long_time_avg_final} is an important result of our paper and
relates the long-time expectation value of an observable to the mean
frequency. The quantity $\mathcal O_{\rm sp}$ is the classical value 
of the observable and the second term is the quantum correction 
within the TWA.

For the pendulum we assume the initial Gaussian distribution 
\begin{equation}
    F_0(\theta, p) = \frac{1}{2\pi d^2}\,e^{-(x^2 + p^2)/(2 d^2)},
    \label{eq:F0_pend}
\end{equation}
where $x = (\theta - \pi)\mod 2\pi$. It is
centered around the saddle point, 
analogous to the Wigner distribution of a mean-field state, 
where the width $d \ll 1$
\footnote{
The quantum Hamiltonian of a pendulum in the $\theta$ basis is $
-(\hbar^2/2) \p^2_\theta + 1 - \cos\theta$. The ground state is
(approximately) a coherent (Gaussian) state around $\theta=0$ with width $d =
\sqrt{\hbar/2}$. When the sign of the potential $\cos\theta$ is suddenly
changed, the state becomes dynamically unstable with the initial Wigner
distribution as in Eq.~\ref{eq:F0_pend}.
}.
Both $H_{\rm pend}$ and $F_0(\theta, p)$ are invariant under the
transformations $p \to -p$ and $\theta \to -\theta$. Thus,
the time-evolved distribution function is also invariant 
and observables $\mathcal O(\theta,p)$ that are odd
functions of either $\theta$ or $p$ have a vanishing expectation value
at all times.  In contrast, observables that are even functions in
both $\theta$ and $p$ can have non-vanishing expectation value. 

As an illustration consider $\mathcal O(\theta,p)=p^2$. Its functional form along the two 
separatrix solutions in Eq.~\ref{eq:pend_sptx} is the same, i.e.,  $[p_{\mathrm
S+}(t)]^2 =[p_{\mathrm S-}(t)]^2$ and, using the indicator functions
$\chi_R(s)$ for the pendulum, we find
\begin{align}
\lim_{t\to \infty} \avg{ p^2(t)} 
 &\sim
  \frac{\avg{\omega_1}_A + \avg{2\,\omega_1}_B + \avg{\omega_1}_C}{2\pi}
  \int_{-\infty}^{\infty}dt\, p^2_{S+}(t).
    \label{eq:pend_p_inf1}
\end{align}
Next we realize that
\begin{align}
\lim_{t\to \infty} \avg{ p^2(t)} 
 &\sim
 \frac{\avg{\varpi}}{2\pi}
  \int_{-\infty}^{\infty}dt\, p^2_{S+}(t) 
  = 
 \frac{8\avg{\varpi}}{2\pi},
    \label{eq:pend_p_inf2}
\end{align}
where we have used Eq.~\ref{eq:pend_sptx} to evaluate the time integral
and defined the ``auxiliary frequency'' $\varpi$ to be $\omega_1$ in region
$A$, $C$ and $2\omega_1$ in region $B$ with average
$\avg{\varpi} =\avg{\omega_1}_A + \avg{2\,\omega_1}_B + \avg{\omega_1}_C$.
From the definition of $\bar f_{0,R}(I_1)$ we also find that
\begin{equation}
\avg{\varpi} \equiv\int_0^\infty d\varpi\, \mathcal \varpi \mathcal F(\varpi),
    \label{}
\end{equation}
where the unit-normalized distribution function
\begin{eqnarray}
    \mathcal F(z) &=& \sum_R \int dI_1 \int_0^{2\pi} \frac{d\varphi_1}{2\pi}\,  f_{0, R}( I_1,\varphi_1)\,\delta(z - \varpi(I_1)) \nonumber 
 \\
 &=&\int_\Omega d\theta dp \, F_0( \theta,p)\,\delta(z - \varpi(\theta,p))
  \label{eq:Fdelta0}
\end{eqnarray}
and $\delta(z)$ is the Dirac delta function.
The second equality shows that the explicit relationship between $(I_1, \varphi_1)$
and $(\theta, p)$ is not required for the analysis.

As shown in App.~\ref{app:pendulum_a} the distribution $\mathcal F(\varpi)$ is
well approximated by a Gaussian when the width $d$ of the initial distribution
$F_0(\theta, p)$ approaches  zero. In fact, the location of
its peak value is
\begin{align}
    \mu\equiv \avg{\varpi} &\sim \frac{2\pi}{\ln(32/(\varkappa d^2))} \ll 1
    \label{eq:mu}
    \end{align}
and its width is
    \begin{align}
    \sigma & \sim \frac{\mu^2} {2\pi\sqrt{1-\varkappa^2}} \ll \mu,
    \label{eq:nu}
\end{align}
where $\varkappa = 0.595\cdots$. Thus, the quantum correction to the long-time 
expectation value of $p^2(t)$ is $1/O(\ln |d|)$.

\section{Time dynamics of relaxation}
\label{sec:relax}

In this section we study the relaxation of an observable to its long-time
expectation value. Observables again depend on only a single angle
$\varphi_1$ and are periodic in $\varphi_1$. We can then write an 
observable in region $R$ as a
Fourier series 
\begin{equation}
     {\mathcal O_R}(\mathbf I, \varphi_1)=
    \sum_{m=-\infty}^\infty \Theta_R( \mathbf I;m) e^{im\varphi_1},
    \label{eq:O_fourier_series}
\end{equation}
with
\begin{equation}
\Theta_R(\mathbf I;m) = 
\int_{0}^{2\pi} \frac{d\varphi_1}{2\pi}
\mathcal O_R(\mathbf I, \varphi_1)\, e^{-i m \varphi_1}.
    \label{eq:O_fourier_def}
\end{equation}
Now, as in Sec.~\ref{sec:long_time_avg}, we transform the integral over $\varphi_1$
into one over time by choosing a reference trajectory with $\varphi_1(0) = 0$
and insert $\varphi_1(t)= \om_1(\mathbf I) t$.
Using Eq.~\ref{eq:Ot} we find
\begin{align}
    \Theta_R( \mathbf I;m) &\sim
    \mathcal O_{\rm sp}\, \delta_{m0}
     + \sum_s  \chi_R(s) 
    e^{-im \alpha_R(s)} \times \\
    &
    \sum_{n=-\infty}^\infty
    \int_{-t_{0,R}(s)}^{T_R-t_{0, R}(s)}\frac{d\tau}{2\pi}\,
    \omega_1  \mathcal D_s(\tau-nT_R)e^{-im\omega_1 \tau} ,
    \nonumber 
\end{align}
where $\delta_{ij}$ is the Kronecker delta, $\alpha_{R}(s) = \omega_1
t_{0, R}(s)$, the integration variable $\tau = t - t_{0,R}(s)$
and we have suppressed the dependence of $\omega_1$ and $T_R$ 
on $\mathbf I$.
Only the $n=0$ term contributes and
\begin{eqnarray}
   \lefteqn{\Theta_R( \mathbf I;m) \sim
    \mathcal O_{\rm sp}\, \delta_{m0} \quad } \label{eq:step}\\
  &&  + \,\sum_s  \chi_R(s) e^{-im \alpha_R(s)} 
   \omega_1  \mathfrak D_s(m\omega_1),
   \nonumber 
\end{eqnarray}
where the Fourier transform 
$
\mathfrak D_s\left(x\right) = 
\int_{-\infty}^\infty dt/(2\pi)\, \mathcal D_s(t) e^{-i x t }$. 
Substituting this expression into Eq.~\ref{eq:O_fourier_series} and
using $\mathcal O_R(t) \equiv \mathcal O_R(\mathbf I, \varphi_1(t))$
Eq.~\ref{eq:def_avg3} becomes
\begin{eqnarray}
    \avg{\mathcal O(t)} &\sim& \mathcal O_{\rm sp} +
    \label{eq:O_t_avg}
   \sum_{m=-\infty}^\infty \sum_{R, s}
        \chi_R(s)e^{-im\alpha_R(s) }\\
       && \quad\quad \times \avg{
        \omega_1  \mathfrak D_s\left(m\omega_1\right)
        e^{i m [\omega_1 t + \varphi_1(0)]  }
        }_R,
   \nonumber 
\end{eqnarray}
where ${\avg \dots}_R$ is the average over $f_{0,R}({\mathbf I},
\bm \varphi)$, the initial distribution restricted to region $R$.
We realize that at long 
times all Fourier terms except the $m=0$ term must go to zero in order
to recover Eq.~\ref{eq:long_time_avg_final}.

We now specialize to the pendulum system. The phases $\alpha_R(s)$ 
are $\alpha_A(S+)= \alpha_C(S-) = \pi$,
$\alpha_B(S-)=\pi/2$ and $\alpha_B(S+) = 3\pi/2$ when $\chi_R(s)$ is nonzero
and, as shown in App.~\ref{app:pendulum_b}, we have
\begin{align}
    \avg{\mathcal O(t)} &\sim \mathcal O_{\rm sp}   \label{eq:pndm_O_t}\\
    \quad &+  \sum_{m=-\infty}^\infty (-1)^m\int_0^\infty d\varpi\, \mathcal F(\varpi) 
  \varpi    \mathfrak D_{S+}(m\varpi) e^{im\varpi t}\,, \nonumber
\end{align}
where, as in Sec.~\ref{sec:long_time_avg}, the auxiliary frequency $\varpi$
is $\omega_1$ in regions $A$, $C$ and $2\omega_1$ in region $B$.
The distribution $\mathcal F(\varpi)$ is well approximated by a Gaussian 
with mean and width given in
Eqs.~\ref{eq:mu} and \ref{eq:nu}, respectively.
The factor 
$\varpi \mathfrak D_{S+}(m\varpi)$ is slowly varying across the
width of $\mathcal F(\varpi)$.
Carrying out the integral over $\varpi$ 
in Eq.~\ref{eq:pndm_O_t} (after extending the lower limit of the integral 
to $-\infty$) gives
\begin{equation}
  \avg{\mathcal O(t)} \sim \mathcal O_{\rm sp} + \!\!
    \sum_{m=-\infty}^\infty (-1)^m  \mu \, \mathfrak D_{S+}(m\mu)
       e^{im\mu t-m^2 \sigma^2 t^2/2}.
       \label{eq:pndm_O_t_final}
\end{equation}
Specifically,  for $\mathcal O(\theta, p) = p^2$
we have 
\begin{equation}
    \avg{p^2(t)} \sim 
    \frac{4\mu}{\pi}
    + 
    \sum_{m=1}^\infty (-1)^m
  \frac{4 m \mu^2  \cos(m\mu  t)}{\sinh(\pi m\mu/2)}e^{-m^2\sigma^2t^2/2},
  \label{eq:p2(t)}
\end{equation}
and the time evolution is a sum of oscillatory functions with damping that is Gaussian in time. The oscillation frequency of each term increases linearly 
with $m$, while simultaneously its damping time, $1/(m\sigma)$, decreases.

\section{Condensate in a double-well potential}
\label{sec:DW}

\begin{figure}
  \begin{center}
    \includegraphics[scale=0.5]{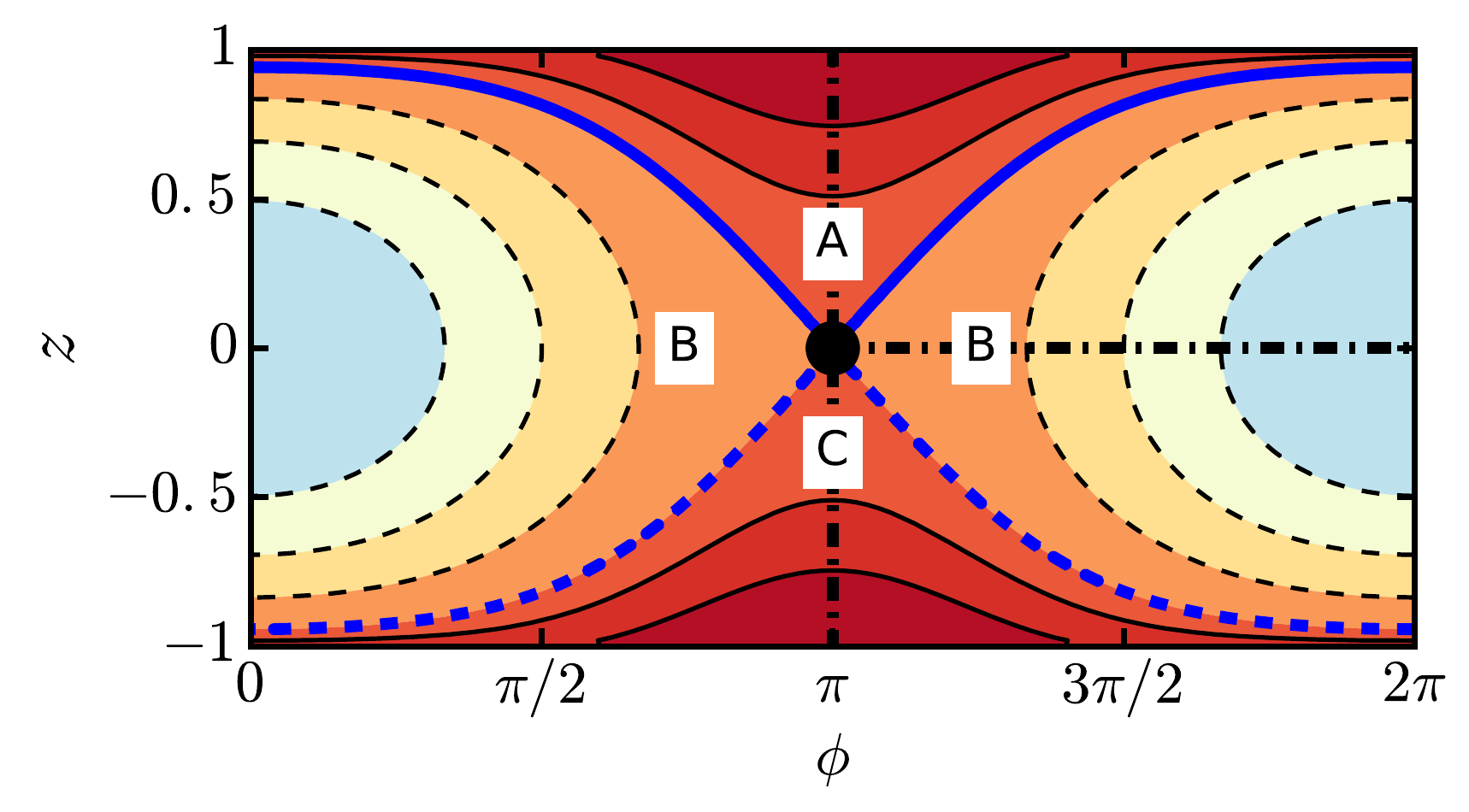}
    \caption{
Equal-energy contours in the phase space $(\phi, z)$ of a condensate in a double-well
potential for $\Lambda =3$. The phase space is equivalent to a sphere,
where the lines $z=1$ and $z=-1$ correspond to the north and south pole,
respectively. Moreover, $(0, z)$ and $(2\pi, z)$ are equivalent. 
Separatrices thick solid blue line ($S+$) and thick dashed blue line 
($S-$) intersect at the saddle point shown by a
solid circle. They divide the phase space into regions $A$, $B$ and $C$.
For each region the thick dashed-dotted black line defines the action-angle 
coordinate $\varphi_1 = 0$.
    }
    \label{fig:DW_phase_space}
  \end{center}
\end{figure}

\begin{figure}
  \begin{center}
    \includegraphics[scale=0.35]{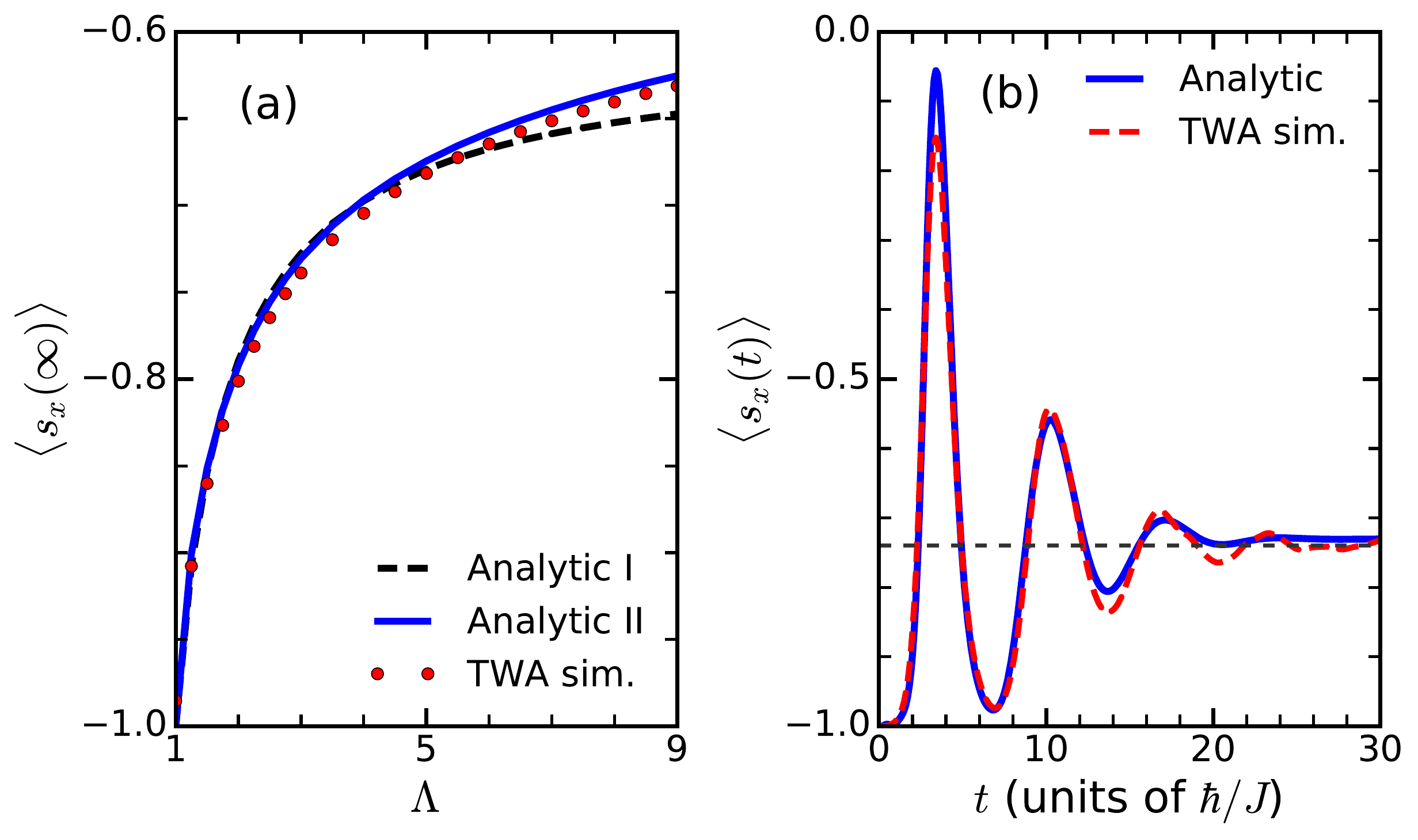}
    \caption{
    Long-time expectation values and time dynamics within the TWA 
    of a Bose-Einstein condensate in a double-well potential following a quench to 
    a dynamically unstable point. Panel (a) shows the long-time expectation
    value of an observable $s_x$ as defined in the text. The dotted black and
    solid blue lines show the analytic result of Eq.~\ref{eq:s_x_inf} with
    mean $\mu\equiv\avg{\varpi}$ given by Eq.~\ref{eq:mu_asym_DW} and with $\mu$ 
    obtained by numerically sampling the initial Wigner distribution, respectively.
    The red circles are values obtained by numerical TWA simulations.  Panel (b)
    shows the time dynamics of $\avg{s_x(t)}$ for $\Lambda=3$. The 
    solid blue line is $\avg{s_x(t)}$ in Eq.~\ref{eq:s_x_t} with 
    $\mu$ and width $\sigma$ obtained by numerically sampling from the
    initial Wigner distribution. The red dashed line is found from 
    numerical TWA simulations.
    For both panels the number of particles $N=1000$.
    \label{fig:DW_x_plots}
}
  \end{center}
\end{figure}

A Bose-Einstein condensate in a weakly-coupled double-well
potential displays Josephson oscillations and macroscopic
self-trapping \cite{smerzi_quantum_1997,raghavan_coherent_1999,
leggett_bose-einstein_2001,albiez_direct_2005,zibold_classical_2010}.
These phenomena are adequately described by a mean-field
approximation. Moreover, dynamical instabilities, where
quantum effects become important, have also been studied
\cite{anglin_dynamics_2001,chuchem_quantum_2010}.

A BEC in a symmetric double-well potential is well
described by assuming that only two  modes $\Psi_1(\vec r)$ and
$\Psi_2(\vec r)$ are occupied, one for each well.  In the mean-field
description the time-dependent order parameter or condensate wavefunction 
is $\psi_1( t)\Psi_1(\vec r)+ \psi_2(t) \Psi_2(\vec r)$ with complex-valued
amplitudes $\psi_j(t)$.  The real and imaginary parts of $\psi_j(t)$
form two pairs of canonical coordinates. Hence, the system has a
four-dimensional phase space. Its classical Hamiltonian is
\begin{equation}
     H_{\mathrm{dw}} =  
       - J(  \psi_1\psi_2^*+ \psi^*_1 \psi_2 ) +
        \frac{U}{2}  (|\psi_1|^4 + |\psi_2|^4)
       \,,
         \label{eq:H_JJ}
\end{equation}
where $U$ and $J > 0$ are the on-site interaction
and tunneling energies, respectively \cite{smerzi_quantum_1997}. 
The total number of atoms $N=|\psi_1|^2 + |\psi_2|^2$ and  
energy $\cal E$ are conserved, making the system is integrable.
We note that the underlying quantum Hamiltonian is solvable 
by the Bethe ansatz \cite{links_algebraic_2003}.

Following the literature it is convenient to introduce $\psi_j( t)=
\sqrt{N_j(t)}e^{i\theta_j(t)}$, where $N_j$ is the number of atoms in and
$\theta_j$ is the phase of the condensate in the $j$-th well \cite{smerzi_quantum_1997}.  We can
then express Eq.~\ref{eq:H_JJ} in terms of the fractional population
difference $z = (N_1 - N_2)/N$ and phase difference  $\phi = \theta_1 -
\theta_2$, where $\phi \in [0, 2\pi]$ and $\phi =0,2\pi$ 
are identical. In fact, we have $H_{\rm dw}= NJ\times h_{\rm dw}(\phi,z)$,
where $h_{\rm dw}(\phi,z)$ is the ``single-atom'' Hamiltonian that depends
on the effective $N$-dependent coupling strength $\Lambda=UN/(2J)$
and is given by
\begin{equation}
    h_{\rm dw}(\phi,z) = 
    \frac{\Lambda z^2}{2} - \sqrt{1-z^2}\cos\phi.
    \label{eq:h_dw}
\end{equation}

The Hamiltonian $h_{\rm dw}(\phi, z)$
has a single minimum located at
$(\phi,z)=(0,0)$ for $\Lambda > 0$. 
For $\Lambda > 1$ the Hamiltonian has a saddle point
located at $(\phi,z)=(\pi,0)$.
Near the saddle
point $h_{\rm dw}(\phi,z) \sim  1+ [(\Lambda-1) z^2- (\phi-\pi)^2]/2$. 
Figure \ref{fig:DW_phase_space} shows equal-energy
contours of $h_{\rm dw}(\phi,z)$ in the two-dimensional phase space
$(\phi,z)$ for $\Lambda > 1$.
Two separatrices $S+$ and $S-$ divide the phase space
into regions $A$, $B$ and $C$. Similar to the pendulum, in region $A$ and $C$
the motion is rotational while in region $B$ it is librational.
Explicit expressions for rotation and libration trajectories are 
given in App.~\ref{app:DW}.
On each separatrix  we consider a trajectory $(\phi_s(t), z_s(t))$ that
only varies significantly around $t=0$ and for which $|z(t)|$ has a maximum
at $t=0$. Along these trajectories
\begin{equation}
    z_{S\pm}(t) =  \pm\frac{2\sqrt{\Lambda -1}}
    {\Lambda} \sech\left( \sqrt{\Lambda-1}t \right).
    \label{eq:z_S}
\end{equation}
The corresponding $\phi_{S\pm}(t)$ can be calculated 
by solving $h_{\rm dw}(\phi_{S\pm}(t), z_{\pm}(t)) = 1$.

We now consider the dynamics of a (zero-temperature) condensate with $N$
atoms prepared at the saddle point within the TWA.
We assume that the initial state is 
$(\psi_1, \psi_2) = (\sqrt{N/2},
-\sqrt{N/2})$ 
with corresponding Wigner distribution 
\begin{equation}
   F_0(\psi_i,\psi^*_i) = \frac{4}{\pi^2} 
   e^{-2|\psi_1-\sqrt{N/2}|^2 - 2|\psi_2+\sqrt{N/2}|^2},
  \label{eq:f0_DW}
\end{equation}
where $i \in \{1, 2\}$ and the probability measure 
is $ \prod_i d\psi_i^*\, d\psi_i $.
The distribution $F_0(\psi_i,\psi^*_i)$ corresponds to 
the Wigner transform of a product of coherent states,
one in each of the two modes with mean atom number $N/2$ and 
a relative phase of $\pi$.

Observables have a natural interpretation as spin operators when 
we represent the phase-space $(\phi, z)$ as
a sphere with polar angle $\vartheta =\arccos(z)$ and 
azimuthal angle $\phi$. Hence, observable $z$
corresponds to $s_z$, the $z$ component of the unit ``spin'' $\vec s$. The other
spin components are $s_x = \sin\vartheta \cos\phi = \sqrt{1-z^2}\cos\phi$
and $s_y =\sin\vartheta \sin\phi = \sqrt{1-z^2}\sin\phi$. As in the
pendulum case, observables that are odd functions of $\phi$ or $z$
have vanishing expectation values for all times. Thus, $\avg{s_z(t)}
= \avg{s_y(t)} =0$, but $\avg{s_x(t)}$ is non-vanishing.
Using Eq.~\ref{eq:h_dw}, we find that $s_{x}
= \Lambda z^2/2 - 1$ on the separatrices.

Now we evaluate the long-time limit and time dynamics of 
$\avg{s_x(t)}$. The indicator functions $\chi_R(s)$ are $\chi_A(S+)=1$, 
$\chi_B(S+)=1$, $\chi_B(S-)=1$, $\chi_C(S+)=1$ and zero otherwise. Then 
using Eqs.~\ref{eq:long_time_avg_final}, \ref{eq:z_S} and following the
derivation in Sec.~\ref{sec:long_time_avg} we find 
\begin{equation}
    \lim_{t \to \infty} \avg{s_x(t)} \equiv \avg{s_x(\infty)}
 \sim -1 + \frac{2\sqrt{\Lambda-1}}{\pi\Lambda} \avg{\varpi},
    \label{eq:s_x_inf}
\end{equation} 
where the auxiliary frequency $\varpi$ is $\omega_1$ in regions $A$, $C$ and
$2\omega_1$ in region $B$.
The time evolution of $\avg{s_x(t)}$ is found by repeating the
steps in Sec.~\ref{sec:relax}.   Details are given in App.~\ref{app:DW},
where we find that the asymptotic expression of $\avg{s_x(t)}$ is again given by
Eq.~\ref{eq:pndm_O_t}, with a distribution function
$\mathcal F(\varpi)$ that is a well approximated by a narrow Gaussian with 
mean $\mu=\avg{\varpi}$ and width $\sigma\ll \mu$ that depend on $\Lambda$ and $N$.
Then Eq.~\ref{eq:pndm_O_t_final} holds and 
\begin{align}
    \label{eq:s_x_t}
    \avg{s_x(t)} &\sim  \avg{s_x(\infty)} + \\
    &\sum_{m=1}^\infty (-1)^m \frac{2 m \mu^2\cos(m\mu t)}
    {\Lambda \sinh[m \mu \pi/ (2 \sqrt{\Lambda-1})]}
    e^{-m^2\sigma^2 t^2/2} \nonumber.
\end{align} 
It is important to note that, as shown in App.~\ref{app:DW_a}, 
for large $N$  
the mean
$\mu$ is $O(1/\ln N)$ and the width is $O[1/(\ln N)^2]$.
Thus, the quantum correction to the long-time value of $\avg{s_x(t)}$ is
$O(1/\ln N)$. Quantitative analytical expressions for $\mu$ and $\sigma$
have only been found for $\Lambda-1\ll 1$.

Figures \ref{fig:DW_x_plots}(a) and (b) show the long-time
expectation value Eq.~\ref{eq:s_x_inf} as a function of $\Lambda$ and
Eq.~\ref{eq:s_x_t} as a function of time, respectively.  In addition, the
figures show good agreement with numerical TWA results. In the numerical
implementation of TWA we sample from the initial distribution $F_0(\psi_i,
\psi_i^*)$, propagate the classical equations of motion and compute the
expectation value of an observable by averaging over the sample.

\section{Spinor BEC within the single-mode approximation}
\label{sec:spinor}

A trapped spin-1 (spinor) Bose-Einstein condensate 
is well-described by a single spatial mode for its three
magnetic sublevels \cite{law_quantum_1998,zhou_spin_2001,zhang_coherent_2005}. This
single-mode approximation (SMA) is valid when the spin healing length,
the length scale over which the spin populations of the condensate
can change significantly, is larger than the condensate size.
The mean-field theory within SMA has turned out to adequately
describe atomic spinor experiments with strong spatial confinement
\cite{chang_coherent_2005,black_spinor_2007,liu_quantum_2009,bookjans_quantum_2011}.
Quenches to dynamical instability, where quantum
effects need to be treated, have also been studied experimentally
\cite{hamley_spin-nematic_2012,gerving_non-equilibrium_2012}. 

\begin{figure}
    \begin{center}
        \includegraphics[scale=0.5]{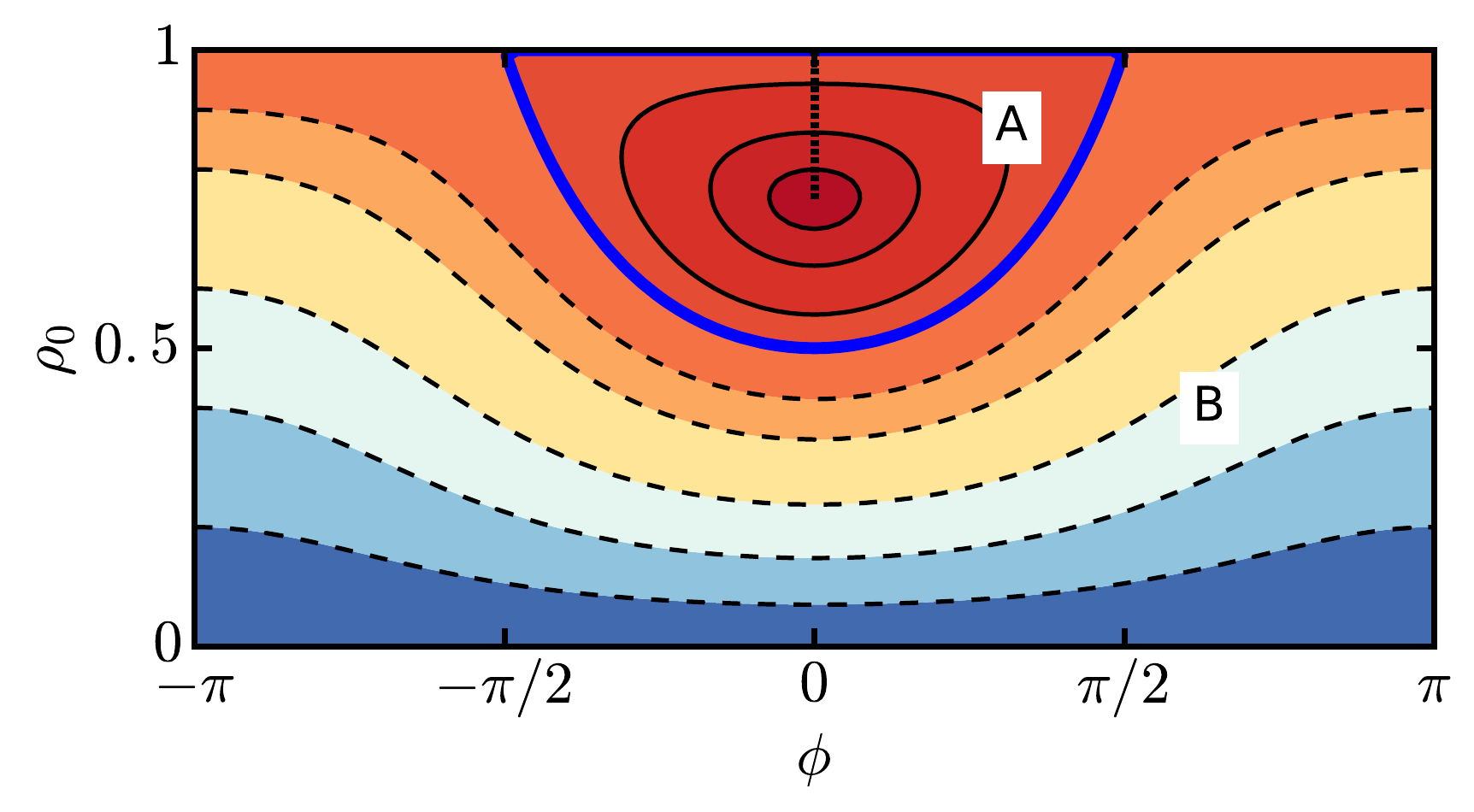}
    \end{center}
    \caption{ 
Equal-energy contours in the phase space $(\phi,\rho_0)$ of an
antiferromagnetic spin-1 condensate in the single-mode and mean-field 
approximations. The magnetization $M=0$,  $q=-1$ and $c=1$.  The phase
space is geometrically equivalent to a sphere as the edges $\phi=-\pi$
and $\pi$ are equivalent and the lines $\rho_0 =1$ and $\rho_0 =0$ are
identified to the north and south pole, respectively. The 
thick solid blue line is the separatrix ($S$) that divides the phase space into regions
 $A$ and $B$. The saddle point is located
at the north pole $\rho_0=1$.  
(Note that the planar projection of
the sphere incorrectly suggests that this point is a
line segment.) In region $A$ the action-angle coordinate
$\varphi_1$ is zero along the black dotted line,
while in region $B$ it is zero on $\phi= \pm\pi$.
}
\label{fig:spinorphase}
\end{figure}

The mean-field wavefunction of the spinor BEC in the SMA is the vector
$\vec\Psi(\vec r, t) = \left(\psi_{-1}(t),\psi_{0}(t),\psi_{+1}(t)\right)^T
\Phi(\vec r)$, where $\psi_j(t)$ is the complex amplitude of the $j$-th
magnetic sublevel along the  external magnetic
field and $\Phi(\vec r)$ is the time-independent unit-normalized spatial
mode. The phase space spanned by the $\psi_j(t)$ has six dimensions and
the system has three mutually commuting conserved quantities, namely energy,
total atom number $N=\sum_j |\psi_j(t)|^2$, and magnetization 
$M= \sum_j j |\psi_j(t)|^2$.  Thus, the system is integrable.
We note that the underlying quantum few-mode 
Hamiltonian is solvable by the Bethe ansatz \cite{bogoliubov_spinor_2006,lamacraft_spin-1_2011}.

It is convenient to write $\psi_j(t)=\sqrt{N_j(t)} e^{i\theta_j(t)}$,
where $N_j$ and $\theta_j$ are the number of atoms in and the condensate
phase of sublevel $j$, respectively. Non-trivial dynamics of the
spinor system occurs in a reduced two-dimensional space $\Omega_{2D}$
with coordinates $\phi$ and $\rho_0$, for a fixed $N$ and $M$.  Here,
$\phi = \theta_1 + \theta_{-1} - 2\theta_0$, where $\phi \in [-\pi, \pi]$
and $\phi = \pm \pi$ are identical; and $\rho_0=N_0/N$ is the fraction of
atoms in the $j=0$ sublevel.  In these coordinates the system obeys  the
``single-particle'' classical Hamiltonian \cite{zhang_coherent_2005}
\begin{align}
  h_{\rm spin}(\phi,\rho_0)  &= c \rho_0 \left( (1-\rho_0) 
+ \sqrt{(1-\rho_0)^2 -m^2}\cos\phi \right) \nonumber\\
            &\quad\quad+ q (1-\rho_0) ,
  \label{eq:H_spinor}
\end{align}
where the coupling strength $c = g_2 N \int d^3r\, |\Phi(\vec r)|^4$ is $N$ dependent, 
$g_2$ is the
spin-changing atom-atom interaction strength,
the term $q(1-\rho_0)$ describes atomic level shifts with controllable strength $q$
(in essence due to the quadratic Zeeman interaction)
and the conserved unit-magnetization
$m = M/N$.

Here, we will only consider a condensate with antiferromagnetic $c>0$
interactions and assume $m=0$.  Figure \ref{fig:spinorphase} shows
equal-energy contours of $h_{\rm spin}(\phi,\rho_0)$ for a representative
$q$ in $(-2c, 0)$.  The Hamiltonian then has a saddle point at the
north pole $\rho_0=1$ and 
$h_{\rm spin}(\phi,\rho_0) \sim (1-\rho_0) \{c(1+\cos\phi)+q\}$
with a linear energy dependence for small positive $1-\rho_0$.
The slope, given in  
$\{\cdots\}$, changes sign twice when $\phi$ goes from 0 to $2\pi$. 
Unlike the pendulum and double-well systems,
there is only one separatrix $S$, which divides the phase space into
regions $A$ and $B$ with rotation and bounded motion, respectively.
The expression for $\rho_0(t)$ along a general trajectory is given in 
App.~\ref{app:spinor}.
The solution along the separatrix that is symmetric about $t=0$ is 
\begin{equation}
    \rho_{0, S}(t) = 1 - (1-y_{1, S}) \sech^2(\Om t),
    \label{eq:spinor_sptx}
\end{equation}
where $y_{1, S} = |q|/(2 c)$ and $\Om = \sqrt{2|q|c(1-y_{1, S})}$. 
By solving $h_{\rm spin}(\phi_S(t),\rho_{0,S}(t)) = 0$ the corresponding $\phi_S(t)$ can be found.

We prepare the system in the mean-field ground state for $q > 0$,
i.e.,  $\rho_0=1$ or equivalently 
$(\psi_{+1}, \psi_{0}, \psi_{-1}) = (0, \sqrt{N}, 0)$. 
The initial Wigner distribution is 
\begin{equation}
   F_0(\psi_j, \psi^*_j) = \frac{8}{\pi^3} e^{- 2|\psi_{-1}|^2-2|\psi_{0}-\sqrt{N}|^2 
        - 2|\psi_{+1}|^2}\,,
        \label{eq:f0_spinor}
\end{equation}
where $j \in \{+1, 0, -1 \}$, corresponding to a coherent state 
for sublevel $j= 0$ with a mean atom number $N$ and zero phase
and vacuum states for sublevels $j = \pm 1$.
The probability measure for the distribution is $\prod_j d\psi_j^* d\psi_j$.

The parameter $q$ is then quenched to a value between $-2c$ and $0$ 
at time $t=0$ and the system becomes dynamically unstable. 
Using Eq.~\ref{eq:long_time_avg_final} with two contributing regions and 
one separatrix, the average  $\avg{\rho_0(t)}$ long
after the quench is given by
\begin{align}
	\lim_{t \to \infty} \avg{\rho_0(t)} \equiv
    \avg{\rho_0(\infty)}  &\sim 1- \avg{\varpi} \frac{1-y_{1,S}}{\pi\Omega},
    \label{eq:avg_rho0_inf_2}
\end{align}
where we used the indicator functions  $\chi_A(S) = \chi_B(S) =
1$ and defined auxiliary frequency $\varpi$ that is now $\omega_1$ in both
regions with average $\avg{\varpi}=\avg{\omega_1}_A+\avg{\omega_1}_B$.
In App.~\ref{app:spinor_a} we show that
$\avg{\varpi} \sim 2\pi \Omega/\ln(16 N)$.
The quantum correction to the long-time value 
of $\avg{\rho_0(t)}$ is, again, $O(1/\ln N)$.

\begin{figure}
  \begin{center}
    \includegraphics[scale=0.35]{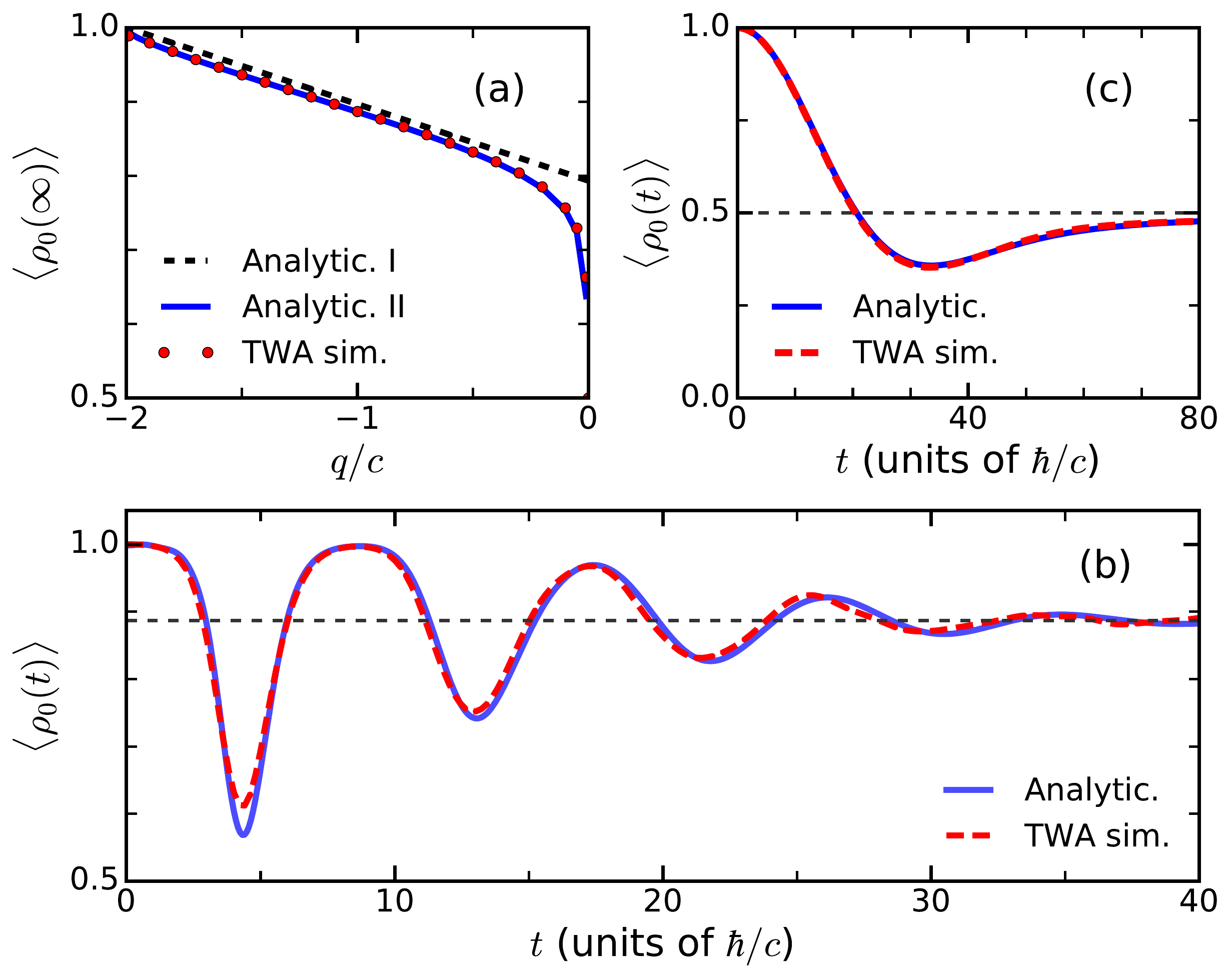}
  \end{center}
  \caption{
Long-time expectation values and time dynamics of a
spin-1 BEC in SMA and TWA after an initial (polar) state with all atoms
in spin projection zero is quenched to a dynamically unstable point with $q < 0$.
The number of atoms $N=1000$.
Panel (a) shows the long-time expectation value
of the fraction of atoms in spin projection zero, 
$\avg{\rho_0(\infty)}$, as a function of $q/c$.
The dashed black line and solid blue curve follow from Eq.~\ref{eq:avg_rho0_inf_2}
with mean $\mu \equiv\avg{\varpi}$ given by our analytical result and
a numerical value as determined from sampling the initial Wigner
distribution, respectively.  
Numerical TWA simulations correspond to the red circles.
Panel (b) shows the time evolution of $\avg{\rho_0(t)}$ for $q/c=-1$.
The solid blue and dashed red curve are obtained from Eq.~\ref{eq:rho0(t)} and numerical TWA simulations, respectively. 
For the solid blue line the mean $\mu$ and width $\sigma$ is obtained by numerical sampling 
the initial Wigner distribution.
Finally, panel (c) shows the time evolution of $\avg{\rho_0(t)}$
for the special case where $q/c =0$. The solid blue curve corresponds
to Eq.~\ref{eq:avg_rh0_t_q_0}, while the nearly-indistinguishable 
dashed red curve is 
from numerical TWA simulations. 
The horizontal dashed lines in panels (b) and (c) 
are the long-time values.
}
  \label{fig:spinor_plot}
\end{figure}

Figure
\ref{fig:spinor_plot}(a) shows $\avg{\rho_0(\infty)}$ as a function
of $q/c$ for $q \in (-2c, 0)$ and fixed atom number $N=1000$. The
analytical expression of $\avg{\rho_0(\infty)}$
with $\varpi = 2\pi \Omega/\ln(16 N)$
gives a straight line. 
The figure also shows the predictions from numerical
TWA for the same parameters. For small negative $q$ the two curves differ
appreciably. We can reproduce the numerical TWA results when we replace
$\avg{\varpi}$ in Eq.~\ref{eq:avg_rho0_inf_2} by its numerical value as
obtained from sampling the initial Wigner distribution.  For $|q|/c$
much smaller than the scale of our figure, however, the $\avg{\rho_0(\infty)}$
from the numerical TWA and that based on computing $\avg{\varpi}$
from sampling still differ.  We will return to this issue later on in
this section.

The time evolution of  $\avg{\rho_0(t)}$ is again calculated
from  Eq.~\ref{eq:O_t_avg}.  The dominant contribution to the
expectation value is from the trajectories 
with the action-angle coordinate $\varphi_1(0)\approx 0$
(See App.~\ref{app:spinor_b} for a formal justification.) Hence, we can
set $\varphi_1(0)=0$ and with $\alpha_A(S) = \alpha_B(S) = \pi$  find
\begin{equation}
    \avg{\mathcal \rho_0(t)} \sim 1+ \sum_{R=A,B}
    \sum_{m=-\infty}^\infty (-1)^m\avg{\omega_1 \mathfrak 
    D_S(m\omega_1) e^{im \omega_1 t}}_R,
    \label{}
\end{equation}
where $\mathfrak D_S(x)$ is the Fourier transform of $\rho_{0, S}(t)$.
As in the previous section, we define the distribution function 
$\mathcal F(\varpi)$ with $\varpi=\omega_1$ in both regions. It
is approximately Gaussian with mean $\mu=\avg{\varpi}$
and width $\sigma\ll\mu$ (see App.~\ref{app:spinor}). 
Then, in a manner similar to that used to find Eq.~\ref{eq:p2(t)}, we derive
\begin{align}
  \label{eq:rho0(t)}
  \avg{\rho_0(t)} &\sim 
  \avg{\rho_0(\infty)}
  - (1-y_{1,S}) \\
  &\quad \times \sum_{m=1}^\infty (-1)^m 
  \frac{m \mu^2\cos( m\mu t)}{\Omega^2\sinh[m\mu \pi/ (2\Omega)]}e^{-m^2\sigma^2 t^2/2}\,.
    \nonumber
\end{align}
Figure \ref{fig:spinor_plot}(b) shows the typical behavior of
$\avg{\rho_0(t)}$ as a function of time. 
For long times the evolution is a
damped sinusoid oscillating around its asymptotic value, as only one term
in the sum significantly contributes. For shorter times the evolution
is more complex and multiple terms are important. 
The numerical TWA
simulations are in good agreement with our analytical expression.

At $q=0$ the Hamiltonian $h_{\rm spin}(\phi,\rho_0)$ has a degenerate
line of saddle points along $\phi=\pi$, instead of a single saddle point.
The system is then critical and the formalism described so far can not 
be applied.  Nevertheless, we show in App.~\ref{app:spinor_c} that
\begin{equation}
    \avg{\rho_0(t)} \sim 1- \, \alpha t\, F(\alpha t),
  \label{eq:avg_rh0_t_q_0}
\end{equation}
where $\alpha = c \,\sqrt{2/N}$ and $F(x)$ is the Dawson integral \cite{dlmf}.
Figure \ref{fig:spinor_plot}(c) shows this evolution as a function of time. The motion seems 
overdamped with little oscillatory behavior.
Agreement with  TWA simulation results is very good.

\section{Conclusions and outlook}
\label{sec:conclusion}

We have analytically studied the time dynamics of two integrable bosonic
systems within the truncated Wigner approximation (TWA) when they become
dynamically unstable after a quench in a system parameter.
The initial Wigner distribution is then centered around a saddle point.
We considered a Bose-Einstein condensate (BEC)
in a symmetric double-well potential and an antiferromagnetic spinor BEC
in the single-mode approximation. Using action-angle variables 
and the concept of phase-space mixing we derived the long-time
expectation value of observables, Eq.~\ref{eq:long_time_avg_final}.
We also derived the relaxation dynamics of the expectation value as
given in Eq.~\ref{eq:O_t_avg}. We used a simple pendulum as a guide 
for these derivations.

The time dynamics of the expectation value of an observable is determined
by the distribution of frequency $\omega_1$ of the classical, periodic
trajectories. The evaluation of the time dynamics simplified due to
the symmetries of the Hamiltonian and the initial Wigner distribution.
These symmetries also motivated the definition of an auxiliary frequency $\varpi$,
which has a simple relationship to $\omega_1$.  
For the two bosonic systems when the initial state is a coherent 
state of $N$ atoms the mean of $\varpi$ is $O(1/\ln N)$.
Hence, the deviation of the long-time expectation value from its
classical value at the saddle point is $O(1/\ln N)$. The mean
determines the typical time scale of the oscillations in 
the time evolution.
The width of $\varpi$ is $O[1/(\ln N)^2]$ and determines
the relaxation rate. Furthermore, we obtained their explicit 
dependence on external parameters.

Although we only considered a representative observable for each system,
the time dynamics of observables that quantify (condensate) phase or
squeezing can be readily computed using our formalism. 
Our results are also directly applicable to other integrable systems with 
a single saddle point in phase space, such as a (anti-)~ferromagnetic
spinor BEC with nonzero magnetization and a BEC in an asymmetric
double-well potential.  The formalism can be generalized to integrable
Hamiltonians with multiple saddle points, for example the
Lipkin-Meshkov-Glick model \cite{ribeiro_exact_2008}.

We give a brief outlook on the full quantum dynamics of our two
bosonic systems and its comparison with the TWA. The Hilbert space of
their underlying few-mode quantum Hamiltonians scales linearly with
$N$ when restricted to fixed values of conserved quantities. Thus,
quenches in these quantum systems can be simulated efficiently
on a classical computer. The eigen-energies near the saddle
point have been studied using the Wentzel-Kramers-Brillouin (WKB)
approximation for a BEC in a double-well potential \cite{nissen_2010}
(and the Jaynes-Cumming model \cite{casati_1995,keeling_2009,babelon_2009}).
The anharmonicity in the energy-level spacing defines the quantum break
time \cite{casati_1995}, which scales as $O(\ln N)$ near the saddle
point \cite{casati_1995,nissen_2010}. In fact, we find (not discussed
here in detail) that the TWA diverges from the quantum dynamics after the
first oscillation consistent with this quantum break time. A detailed
study will be the subject of a future publication.

\begin{acknowledgments}

This work has been supported by the National Science Foundation Grant No.~PHY-1506343.
R.~M. acknowledges useful discussions with W.~Sengupta on phase-space 
mixing in classical systems.

\end{acknowledgments}

\appendix

\section{Pendulum}
\label{app:pendulum}

The simple pendulum is used throughout to illustrate our derivation of 
dynamics and long-time expectation values for few-mode integrable systems. 
In this appendix we derive  results specific to the pendulum. 
Its Hamiltonian is given in Eq.~\ref{eq:Hpend} with canonical coordinates 
$\theta$ and $p$ satisfying $\{\theta, p \} =1$, where $\{\cdot, \cdot \}$
is the Poisson bracket.

First, librational trajectories $(\theta_B(t), p_B(t))$ in phase-space
region $B$ are \cite{dlmf}
\begin{align}
  \sin\left( \theta_B(t)/2 \right) 
  &= k \, \sn\left(t+t_0, k \right),\\
  p_B(t) &= 2 k \cn\left(t+t_0, k\right),
  \label{eq:pend_EOM_pl}
\end{align}
where the modulus $k = \sqrt{\mathcal E/2}$, $\mathcal E$
is the energy of the trajectory and time $t_0$ depends on the initial condition.
Secondly, rotational trajectories $(\theta_R(t), p_R(t))$ in  regions $R=A$ and $C$  are
\begin{align}
  \sin\left(\theta_R(t)/2\right) 
  &=  \pm\sn\left((t+t_0)/k, k\right), \\
  p_R(t) &= 
  \pm 2 /k \,
  \dn
  \left((t+t_0)/k, k\right),
  \label{eq:pend_EOM_pr}
\end{align}
where $k = \sqrt{2/\mathcal E}$. The $+$ and $-$ sign correspond
to region $A$ and $C$, respectively.  
The functions $\sn(z, k)$, $\cn(z, k)$ and $\dn(z, k)$ are Jacobi 
elliptic functions \cite{dlmf}. Finally, on the separatrices $\mathcal
E =2$ with trajectories $(\theta_{S\pm}(t), p_{S\pm}(t))$ given by
Eq.~\ref{eq:pend_sptx}.

\subsection{Distribution function $\mathcal F(\varpi)$}
\label{app:pendulum_a}

In this section we calculate the
distribution function
\begin{eqnarray}
    \mathcal F(z) 
 &=&\int_\Omega d\theta dp \, F_0( \theta,p)\,\delta(z - \varpi(\theta,p)),
 \label{eq:Fdelta_app}  
\end{eqnarray}
as defined in Eq.~\ref{eq:Fdelta0}, as well as its mean and width.  Here,
the integral is over the whole phase space $\Omega$ and the initial
Gaussian distribution $F_0(\theta, p)$, given in Eq.~\ref{eq:F0_pend},
has a width $d$ along both $\theta$ and $p$.  The auxiliary frequency
$\varpi(\theta,p)=\omega_1=\pi/[k K(k)]$ in regions $A$ and $C$ and
$\varpi(\theta,p) = 2\omega_1= \pi/K(k)$ in region $B$,
where $K(k)$
is the complete elliptic integral of the first kind with modulus $k\in
[0,1]$ \cite{dlmf}.

Near the saddle point the energy
$\mathcal E \sim 2 + (p^2 -q^2)/2$, where $q = (\theta - \pi) \mod 2\pi$.
The relationship between energy and modulus leads to $k^2 \sim 1 -
|p^2 - q^2|/4$ in {\it all} regions. Finally, $\varpi \sim \pi/ K(k)
\sim 2\pi/\ln(64/|p^2-q^2|)$ using the asymptotic expansion $K(k)\sim
\ln(16/k'^2)/2$ around $k=1$ with complementary modulus $k'$ defined
by $k'^2=1-k^2$.

To compute $\cal F(\varpi)$ it is convenient to first introduce
the invertible transformation $\mathcal X(\varpi) = 2k'^2/d^2\sim 32 e^{-2\pi/\varpi}/d^2$. 
The dependence of $\mathcal X(\varpi)$ on $d$ will become 
clear later. We then write
\begin{equation}
  \mathcal F(\varpi) \sim 
  \frac{2 \pi}{\varpi^2 } \mathcal X(\varpi)
    P\left( \mathcal X(\varpi) \right),
    \label{eq:omega_m'_dist}
\end{equation}
as $d\to 0$ with the distribution
\begin{equation}
    P(\mathcal X) = \int_\Omega d\theta dp \, 
    F_0(\theta, p)\,
    \delta\left(\mathcal X -\frac{|p^2 -q^2|}{2 d^2}\right),
  \label{eq:m'_dist}
\end{equation}
and the factor in front of 
$P(\mathcal X(\varpi))$ in the right-hand side of Eq.~\ref{eq:omega_m'_dist} is the Jacobian $d\mathcal X/d\varpi$.

The separatrices divide the neighborhood of the saddle point into four quadrants.
We solve Eq.~\ref{eq:m'_dist} in each quadrant separately.
For the quadrant in region $A$ ($p>0$ and $p>q$)  we change the 
integration variables to $p = d\sqrt{2 \mathcal X} \cosh u$ and $q = d\sqrt{2 \mathcal X} \sinh u$ with $u \in (-\infty, \infty)$.
Similar changes of  variables can be used in the other three
quadrants (noting that two quadrants lie in region $B$). The contribution to $P(\mathcal X)$ from each quadrant turns out to be the same and
we finally find
\begin{equation}
    P(\mathcal X) = 
    \frac{2}{\pi}K_0\left( \mathcal X \right),
    \label{eq:PX}
\end{equation}
which has no explicit dependence on the width $d$, and
$K_0(x)$ is a modified Bessel function \cite{dlmf}.
We then have
\begin{equation}
  \mathcal F(\varpi) \sim  
  \frac{128 e^{-2\pi/\varpi} }{d^2 \varpi^2} 
    K_0\left( \frac{32 e^{-2\pi/\varpi}}{d^2} \right),
  \label{eq:F_pi}
\end{equation}
as $d\to 0$ and for $\varpi \ll 1$.

\begin{figure}
  \begin{center}
    \includegraphics[scale=0.5]{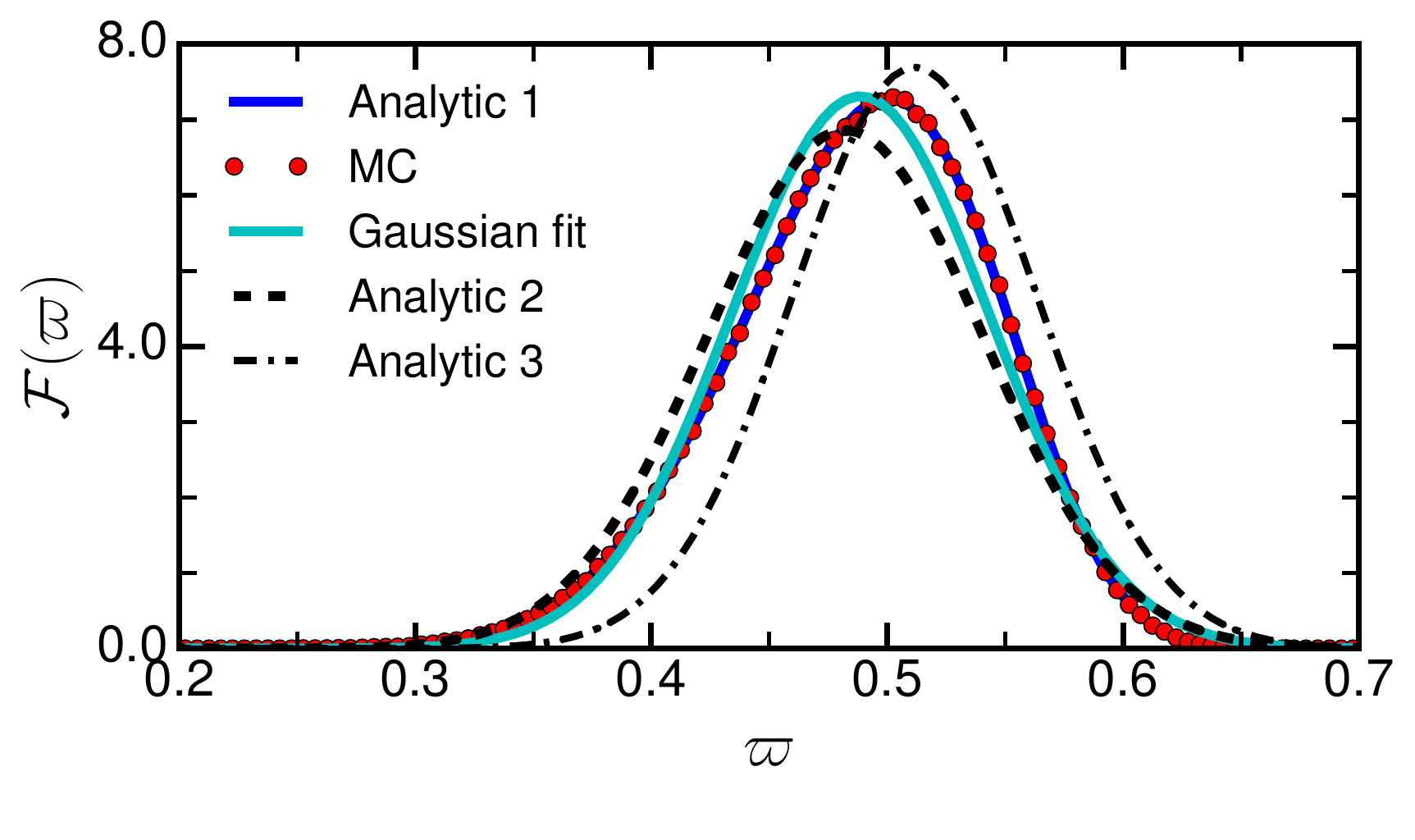}
  \end{center}
  \caption{
Distribution function $\mathcal F(\varpi)$ as a function of the auxiliary 
frequency $\varpi$ for the pendulum with an initial Wigner distribution 
(Eq.~\ref{eq:F0_pend}) 
with width $d=1/20$. 
The blue solid line is the distribution in Eq.~\ref{eq:F_pi}. 
Indistinguishable from this curve is the
$\mathcal F(\varpi)$ shown by red circles, 
which are obtained by numerical Monte-Carlo sampling of the
initial Wigner distribution.
The cyan solid, black dashed, and black dash-dotted lines are Gaussians 
whose mean and standard deviation are given by that of 1) the numerical
distribution, 2) Eqs.~\ref{eq:mu_asym1} and 
\ref{eq:sig_asym1}, and 3) Eqs.~\ref{eq:mu_asym2} and \ref{eq:sig_asym2}, respectively.
  }
  \label{fig:dist_pend}
\end{figure}

Figure~\ref{fig:dist_pend} shows $\mathcal F(\varpi)$ as a function of
$\varpi$ for a single $d$.  We find that $\mathcal F(\varpi)$ is sharply
peaked.  It approaches zero as $ C\, e^{-2\pi/\varpi}/\varpi^3$ when
$\varpi \to 0^+$ and $C$ is a constant.  For $\varpi \gtrsim 1$, where
Eq.~\ref{eq:F_pi} is invalid, either $p$ or $q$ is much greater than $d$
and $F_0(\theta, p)$, hence $\mathcal F(\varpi)$, is exponentially small.
Thus, it is reasonable to approximate $\mathcal F(\varpi)$ by a Gaussian
as shown in Fig.~\ref{fig:dist_pend}.

We now calculate the mean and variance of $\varpi$
using one of two methods. 
The mean $\mu \sim \int_0^\infty d\mathcal X\, P(\mathcal X) 
\varpi({\mathcal X})$ with $\varpi({\mathcal X}) = 2\pi/ \ln[32/(d^2 \mathcal X)]$. 
We then identify the small parameter $\lambda^{-1} = \ln(C/d^2)^{-1}\ll 1$, where 
the constant $C$ will be determined later, and find
\begin{align}
  \mu &\sim
  \frac{2\pi}{\lambda} +
  \frac{2\pi}{\lambda} \sum_{n = 1}^\infty 
        \frac{{\rm E}[\mathcal Y^n]}{\lambda^n},
  \label{eq:mu_asym_series}
\end{align}
with the help of the geometric series. Here, $\mathcal Y = \ln (C \mathcal
X/32)$ and ${\rm E}[\mathcal Y]$ is the expectation value of $\mathcal
Y$ with respect to $P(\mathcal X)$.  For $C = 64 e^\gamma$, where $\gamma$
is the Euler-Mascheroni constant, the expectation value $E[\mathcal Y]
= 0$. Hence,
\begin{equation}
  \mu \sim \frac{2\pi}{\lambda}
            + O(1/\lambda^3),
  \label{eq:mu_asym1}
\end{equation}
Similarly, the variance 
\begin{align}
  \sigma^2 
  &\sim 
  \left( \frac{2\pi}{\lambda} \right)^2
  \biggl[
      \frac{ {\rm E}[\mathcal Y^2] -{\rm E}[\mathcal Y]^2 }{\lambda^2}
  + 
  O(1/\lambda^3)
  \biggr]
  \label{eq:sig_asym_series}
\end{align}
and evaluation of the second moment of $\mathcal Y$ gives
\begin{equation}
  \sigma \sim \frac{\pi^2}{\lambda^2} + O(1/\lambda^3).
  \label{eq:sig_asym1}
\end{equation}
Thus, we find
$\mu = O(1/|\ln d|)$ and $\sigma = O(1/|\ln d|^2)$.  

The second method estimates $\mu$ and $\sigma$ from
 the location of and curvature at the maximum of 
$\mathcal F(\varpi)$ using the fact that the distribution is well 
approximated by a narrow Gaussian. 
We could not find a closed form for maximum of $\mathcal F(\varpi)$.
Instead, we present 
results based on the extremum of $\varpi^2 \mathcal F(\varpi)$.
This only introduces small corrections 
as $\varpi^2 \mathcal F(\varpi) \sim \mu^2 \mathcal F(\varpi)$
over the width of the distribution near $\varpi=\mu$.
After some algebra we find
\begin{equation}
    \mu \sim  \frac{2\pi}{\ln \left(
    32/(\mathcal \varkappa d^2)
    \right)},
  \label{eq:mu_asym2}
\end{equation}
\begin{equation}
  \sigma \sim \sqrt{- \left. \frac{ g(\mathcal X)}
  {d^2 g(\mathcal X)/d\varpi^2}\right\rvert_{\varpi= \mu} 
  }
  = \frac{\mu^2}{2\pi \sqrt{1-\varkappa^2}}
  \label{eq:sig_asym2}
\end{equation}
where $g(\mathcal X)=\varpi^2 \mathcal F(\varpi) = 
4 \mathcal X K_0(\mathcal X) $  and $\varkappa = 0.595\cdots$
is the solution of $dg(\mathcal X)/d\mathcal X=0$.

The estimates of $\mu$ and $\sigma$ obtained by either method
gives the same logarithmic scaling with $d$. The 
numerical prefactors inside the logarithm, however, are different.
Figure \ref{fig:dist_pend} shows  Gaussian distributions with  
the estimated mean and width based on the two methods. Their 
difference from the true ${\cal F}(\varpi)$ vanishes as $d\to 0$.

\subsection{Time dynamics of observables}
\label{app:pendulum_b}

In this subsection we derive the time dynamics of observables for a 
pendulum. That is, we derive Eq.~\ref{eq:pndm_O_t} from Eq.~\ref{eq:O_t_avg}. 
The dependence of the quantity in the bracket $\langle \cdots\rangle_R$ 
in Eq.~\ref{eq:O_t_avg} on the action-angle coordinates is only through
$\omega_1$ and $\varphi_1$. ( This is also true for the other 
two systems studied in the paper.)
Denoting the quantity by ${\cal A}(\omega_1, \varphi_1)$
it  is then convenient to write
\begin{equation}
  \langle {\cal A} \rangle_R =
    \int d\omega_1 \int \frac{d\varphi_1}{2\pi}  {\cal A}(\omega_1, \varphi_1) g_{0, R}(\omega_1, \varphi_1) ,
\end{equation}
where 
\begin{equation}
    g_{0, R}(\omega_1, \varphi_1) = \int_{\mathcal I} d\mathbf I \int_0^{2\pi} 
    \frac{d\bm \varphi'}{2\pi} f_{0, R}(\mathbf I, \bm \varphi) 
    \delta(\omega_1 - \omega_1(\mathbf I))
    \label{eq:g0R}
\end{equation}
and $\bm \varphi' = (\varphi_2, \dots, \varphi_n)$ are
all the angles {\it except} $\varphi_1$. 
(The time dependence of $\cal A$ is suppressed for clarity.)
For the pendulum with its 2D phase space Eq.~\ref{eq:g0R} simplifies to
 $g_{0, R}(\omega_1, \varphi_1) = dI_1/d\omega_1\,f_{0, R}(I_1, \varphi_1)$,
where  $dI_1/d\omega_1$ is the Jacobian of the transformation between $I_1$
and $\omega_1$. 

The function $g_{0, R}(\omega_1, \varphi_1)$ is concentrated
around a few points in the $(\omega_1, \varphi_1)$ space from the
observation that $F_0(\theta, p)$ is localized around the saddle point.
The justification of this approximation is subtle and technical;
it has been relegated to Sec.~\ref{app:pendulum_b_a}.  We find
that 
\begin{align}
g_{0, R}(\omega_1, \varphi_1)  &\approx\left\{
  \begin{array}{l}
    2\pi\,\overline{g}_{0, A}(\omega_1)\, \delta(\varphi_1),\quad\quad\quad\quad \quad R=A,C\\
     \pi\,\overline{g}_{0, B}(\omega_1)
      \left[   \delta(\varphi_1)+ \delta(\varphi_1-\pi) \right], \ R=B
      \end{array} \right.
  \label{eq:g0_pend}
\end{align}
where $\overline{g}_{0, R}(\omega_1) = \int_0^{2\pi}
d\varphi_1/(2\pi) g_{0,R}(\omega_1, \varphi_1)$ is a marginal distribution.

We can now simplify the average and sums on the right-hand side of Eq.~\ref{eq:O_t_avg} into a  single average for  observables that
are even in $\theta$ and $p$. The bump
functions $\mathcal D_{S+}(t)$ and $\mathcal D_{S-}(t)$ are then identical. 
Moreover, the angular dependence of  $g_{0, B}(\omega_1,
\varphi_1)$ implies that $\avg{e^{i m \varphi_1}}_B =0$ when $m$
is odd so that odd Fourier components in region $B$ do not contribute to 
$\avg{\mathcal O(t)}$. 
(For regions $A$ and $C$ both even and odd Fourier components contribute.)
Using these observations, the definition of the auxiliary frequency $\varpi$ and the values of $\alpha_R(s)$, we combine 
the sum over regions and separatrices  into a single sum and  arrive at Eq.~\ref{eq:pndm_O_t}.

\subsubsection{Derivation of Eq.~\ref{eq:g0_pend}}
\label{app:pendulum_b_a}

We give a quantitative argument for Eq.~\ref{eq:g0_pend}. 
In the evaluation of $\cal F(\varpi)$ in 
App.~\ref{app:pendulum_a} we observed that each quadrant in the neighborhood 
of the saddle point contributes equally. In region $A$, 
where $\varpi=\omega_1$,
a comparison of Eq.~\ref{eq:Fdelta0} and the definition of $\overline{g}_{0, A}(\omega_1)$
shows that  $\overline{g}_{0, A}(\omega_1) \propto \mathcal F(\omega_1)$.
Thus,  $g_{0,A}(\omega_1, \varphi_1)$ is localized  
around $\mu = \avg{\varpi}$ with a width $\sigma \ll \mu$ along the $\omega_1$ coordinate.

Next, we define the standard deviation $\Delta_A(\omega_1)$ of $\varphi_1$ 
with respect to the conditional distribution $g_{0,
A}(\omega_1, \varphi_1)/ \overline g_{0, A}(\omega_1)$ 
at each value of $\omega_1$. 
We now estimate $\Delta_A(\omega_1)$ from the momentum spread 
$\Delta p_A=O(d)$ in region $A$, where $d$ is the width of $F_0(\theta, p)$. 
Using Eq.~\ref{eq:pend_EOM_pr} we find 
\begin{equation}
    p_A = \frac{2}{k}
\dn\left( \frac{\varphi_1}{\omega_1 k} + K(k), k \right),
\end{equation}
where $t_0 = kK(k)$, because $p_A$ is 
minimal when $\varphi_1=0$ (see Fig.~\ref{fig:pendulum_phase_space}(a)).
Now we expect the relevant $\varphi_1$ to be small 
and use
the Taylor expansion $\dn(x+K(k), k) = k' + k' k^2 x^2/2 + \cdots$ 
for small $x$  to find
\begin{equation}
    p_A - p_A^{\rm min} \sim 
    k'\left( \frac{\varphi_1}{\omega_1} \right)^2  .
  \label{eq:p_A}
\end{equation}
where $k' = \sqrt{1-k^2}\sim 4 e^{-\pi/\omega_1}$ and $p_A^{\rm min} = 2k'/k$.
Thus,  the width $\Delta_A (\omega_1) \propto
\omega_1\sqrt{\Delta p_A/k'} \propto \omega_1 e^{\pi/(2\omega_1)}\sqrt{d}$.
At first glance, this relation contradicts
the assumption that $\Delta_A(\omega_1)$ is small because $\Delta_A (\omega_1)$ diverges as $\omega_1
\to 0^{+}$.  From Sec.~\ref{app:pendulum_a}, however, we know that  $\mathcal F(\omega_1)$ and, thus, $\overline g_{0, A}(\omega_1)$
go to zero rapidly as $\omega_1 \to 0^+$.  In fact, at the mean
value $\omega_1= \mu$, given in Eq.~\ref{eq:mu_asym2}, we find
$\Delta_A \left(\mu\right) = O(1/|\ln d|) \ll 1$.  Furthermore,
$\Delta_A (\omega_1)$ remains small where $g_{0, A}(\omega_1, \varphi_1)$
is significant as $\sigma\ll\mu$.  Hence, $g_{0, A}(\omega_1, \varphi_1)$
is localized in both $\omega_1$ and $\varphi_1$. 
(The distribution $f_{0,A}(I_1, \varphi_1)$ is not  localized 
in $\varphi_1$ as it does not approach zero as $\omega_1(I_1) \to 0^+$.)  

The nonzero, albeit small, width of $g_{0,R}(I_1, \varphi_1)$ 
in the $\omega_1$ coordinate leads to mixing in $\varphi_1$. 
On the other hand, the distribution over $\omega_1$ is invariant in time.
We can then replace the narrow distribution $g_{0, A}(\omega_1, \varphi_1)$ 
along $\varphi_1$ by a delta function.  That is, $g_{0, A}(\omega_1,
\varphi_1) \approx 2\pi\bar g_{0, A}(\omega_1) \delta(\varphi_1)$.
A similar analysis in regions $B$ and $C$ leads to the other two equations in Eq.~\ref{eq:g0_pend}.

\section{A condensate in a double-well potential}
\label{app:DW}

In this section we derive results pertaining to a two-mode Bose-Einstein condensate 
in a double-well potential. Its ``single-particle'' Hamiltonian $h_{\rm dw}(\phi, z)$  
is defined in  Eq.~\ref{eq:h_dw} and $\{\phi, z \} = 1$.
For $\Lambda>1$ the Hamiltonian has a single saddle point and two
separatrices $S$ divide the phase space into three distinct regions $R=A$, $B$, and $C$.
The solutions to the equations of motion are~\cite{raghavan_coherent_1999} 
\begin{equation}
    z_R(t)= 
    \begin{cases}
    \mathcal C \cn\left( \mathcal C \Lambda (t-t_0)/(2\kappa) , \kappa \right), 
    &R=B\\
     \pm \mathcal C\dn\left( \mathcal C \Lambda (t-t_0)/2, 1/\kappa \right), 
    & R=A,C
    \label{eq:z_t}
  \end{cases}
\end{equation}
where
\begin{eqnarray}
   \mathcal C^2 &=& \frac{2}{\Lambda^2}\left(\mathcal E \Lambda -1 + 
        \sqrt{\Lambda^2 -2\mathcal E \Lambda +1}\right),\\
    \kappa^2&=& \frac{1}{2} + \frac{\mathcal E \Lambda -1}
            {2\sqrt{\Lambda^2 - 2 \mathcal E \Lambda +1}}.
    \label{eq:kappa}
\end{eqnarray}
The ``single-particle'' energy of the trajectory is $\mathcal E$ and
$t_0$ depends on the initial condition. The corresponding $\phi_R(t)$
can be obtained by solving $h(\phi_R(t), z_R(t))= \mathcal E$.  (Note that
Ref.~\cite{raghavan_coherent_1999} misses a factor of $1/2$ in the first
argument of both $\cn(z,k)$ and $\dn(z,k)$.)  Finally, on the separatrices
$\mathcal E = 1$, $\kappa=1$ and $\mathcal C = 2\sqrt{\Lambda -1}/\Lambda$
with solutions $z_{S\pm}(t)$ given by Eq.~\ref{eq:z_S}.

\subsection{Distribution function $\mathcal F(\varpi)$}
\label{app:DW_a}

We now compute the distribution function $\mathcal F(\varpi)$ for a Bose
condensate in a double-well potential. The initial Wigner distribution
Eq.~\ref{eq:f0_DW} is localized around the saddle point 
$(\psi_1, \psi_2) = (\sqrt{N/2}, -\sqrt{N/2})$. It is convenient to 
introduce real coordinates $p_i$ and $q_i$  defined by
$p_1 + i q_1 = \psi_1 - \sqrt{N/2}$ and 
$p_2+iq_2 = \psi_2 + \sqrt{N/2}$. 
In these coordinates the Wigner distribution becomes
\begin{equation}
    F_0(p_i, q_i)
    =\frac{4}{\pi^2}  e^{-2(p_1^2 + q_1^2 + p_2^2 + q_2^2)}\, ,
    \label{eq:random}
\end{equation}
where $i \in \{1,2\}$ and the probability measure is $dp_1 dq_1 dp_2 dq_2$.
Near the saddle point 
\begin{align}
  z &= \sqrt{\frac{2}{N}} (p_1 + p_2) + O(N^{-1}),\\
  \phi &=- \pi +\frac{q_1 + q_2}{\sqrt{N}} +O(N^{-1})
\end{align}
and their substitution into $h_{\rm dw}(\phi, z)$ gives
the energy
\begin{equation}
      \mathcal E = 1 + \frac{1}{N}
      \left[ (\Lambda-1)(p_1 + p_2)^2 - (q_1 + q_2)^2 \right] + O(N^{-3/2})
      \label{eq:EdwAB}
\end{equation}
close to one.

Next, we express the auxiliary frequency $\varpi=\omega_1$ in regions
$A$, $C$ and $2\omega_1$ in region $B$ in terms of coordinates $p_i$ and
$q_i$.  From Eq.~\ref{eq:z_t} and the periodicity of elliptic functions,
it follows that near the separatrix $\varpi \sim \pi \sqrt{\Lambda
-1}/K(k) \sim 2\pi \sqrt{\Lambda-1}/\ln(16/k'^2)$ where $k=\kappa$
in region $B$ and $1/\kappa$ in regions $A,C$.  The modulus $k$ and its
complement $k'$ depend on $\mathcal E$ and thus on the $p_i$ and $q_i$.
With the help of Eqs.~\ref{eq:kappa} and \ref{eq:EdwAB}, we find
\begin{equation}
    \mathcal X \equiv 2\left( \frac{\Lambda-1}{\Lambda} \right)^2 N k'^2 \sim  
  |(\Lambda-1)(p_1 + p_2)^2 - (q_1 + q_2)^2|.
  \label{eq:X_def}
\end{equation}
This choice of $\mathcal X$, in particular its $N$ dependence,
will simplify later derivations. We realize that $\varpi
\sim  2\pi \sqrt{\Lambda-1}/\ln[32N (\Lambda-1)^2/(\mathcal X
\Lambda^2)]$ and $\mathcal X(\varpi) = 32 N (1-\Lambda^{-1})^2
e^{-2\pi\sqrt{\Lambda-1}/\varpi}$.  Thus, we have established a relation
between $\varpi$ and $p_i$, $q_i$ via the variable $\mathcal X$.

\begin{figure}
  \begin{center}
    \includegraphics[scale=0.5]{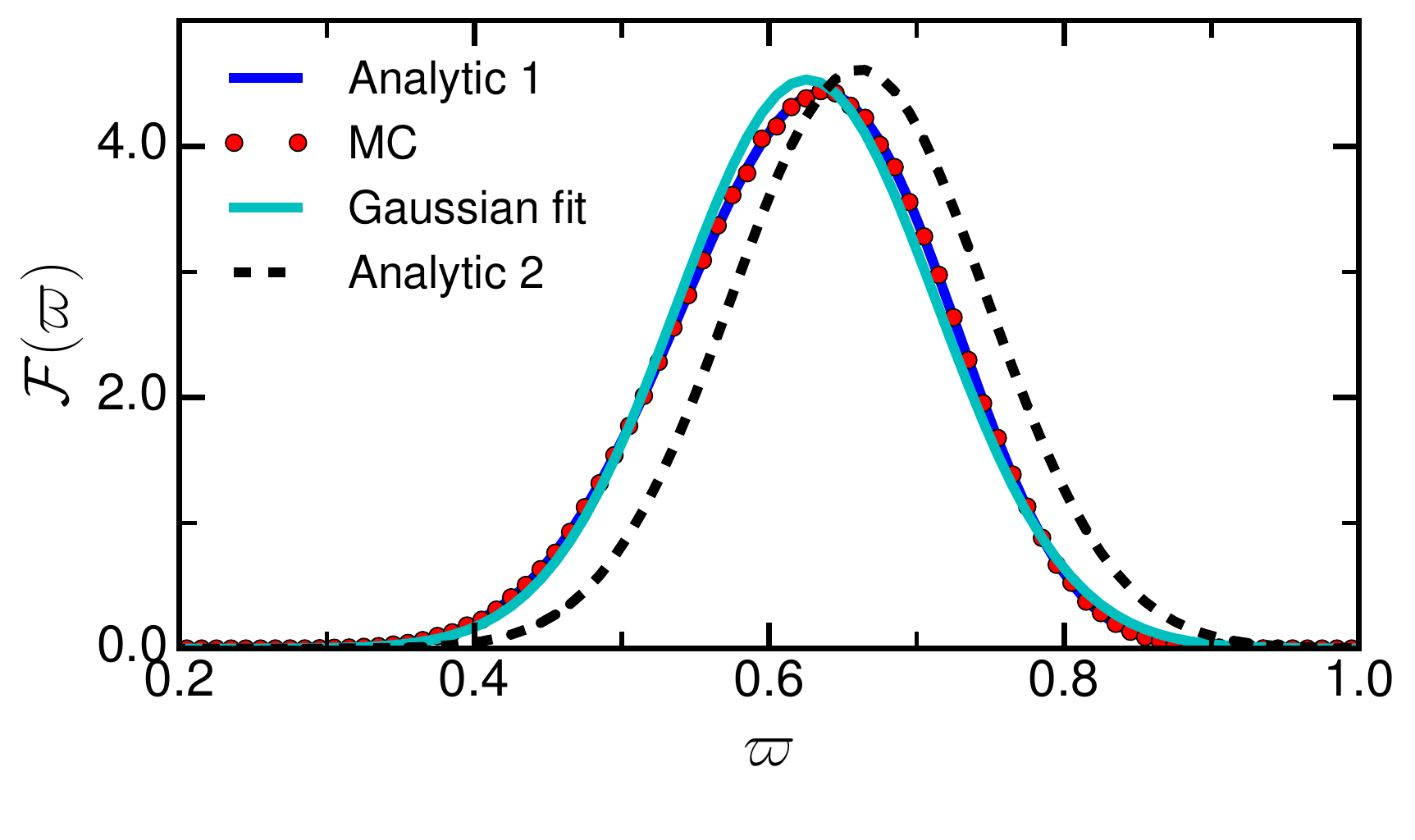}
  \end{center}
  \caption{
Distribution function $\mathcal F(\varpi)$ as a function of  
auxiliary frequency $\varpi$
for a  Bose-Einstein condensate in a double-well potential with $1000$ atoms and 
$\Lambda=2$. The solid blue curve is the distribution in Eq.~\ref{eq:F_pi_DW}.
Indistinguishable from this curve is the $\mathcal F(\varpi)$ shown by red circles,
which is obtained
by Monte Carlo sampling of the initial Wigner distribution.
The cyan solid line is a Gaussian fit to this data. The dashed line is a Gaussian
distribution whose mean and standard deviation is given by
Eqs.~\ref{eq:mu_asym_DW} and \ref{eq:sig_asym_DW}, respectively.
  }
  \label{fig:dist_DW}
\end{figure}

The distribution $\mathcal F(\varpi)$ is then
\begin{equation}
    \mathcal F(\varpi) = \frac{2 \pi \sqrt{\Lambda-1}}
    {\varpi^2} \mathcal X(\varpi)
    {\mathcal P}(\mathcal X(\varpi)),
  \label{eq:F_pi_DW}
\end{equation}
where 
\begin{equation}
  {\mathcal P}(\mathcal X) = \int dp_1 dq_1 dp_2 dq_2\, F_0(p_i, q_i) 
        \,\delta(\mathcal X- \mathcal Z(p_i, q_i )),
  \label{eq:PX_DW}
\end{equation}
with $\mathcal Z(p_i, q_i)$ equal to the right-hand side of
Eq.~\ref{eq:X_def} and the factor  multiplying ${\mathcal P}(\mathcal X)$ 
in Eq.~\ref{eq:F_pi_DW} is the Jacobian $d\mathcal X/d\varpi$.

We simplify the integrals in Eq.~\ref{eq:PX_DW} by changing to ``center
of mass'' and ``relative'' coordinates $P = (p_1 + p_2)/2$, $p = p_1 -
p_2$, $Q = (q_1 + q_2)/2$ and $q = q_1 - q_2$.
We find
\begin{equation}
  {\mathcal P}(\mathcal X) = 
  \frac{4}{\pi}\int_{-\infty}^\infty dP dQ\, 
  e^{-4P^2 - 4Q^2}      
  \delta\left( \mathcal X - 
      4|(\Lambda-1) P^2 - Q^2|
  \right),
  \label{}
\end{equation}
which yields
\begin{equation}
  {\mathcal P}(\mathcal X) = 
  \frac{2}{\pi \sqrt{\Lambda -1}} 
  \cosh\left( \frac{\Lambda-2}{2(\Lambda -1)} \mathcal X\right)
  K_0\left( \frac{\Lambda \mathcal X}{2(\Lambda-1)} \right).
  \label{eq:f_X_Lambda}
\end{equation}
Figure \ref{fig:dist_DW} shows $\mathcal
F(\varpi)$ for $N=1000$ and $\Lambda = 2$. It is evident from the figure
that $\mathcal F(\varpi)$ is well approximated by a Gaussian distribution.
The mean $\mu$ and width $\sigma$ of $\mathcal F(\varpi)$ can be computed from
Eqs.~\ref{eq:mu_asym_series} and \ref{eq:sig_asym_series}, respectively,
with $\lambda = \ln[32 N (1-\Lambda^{-1})^2/\Lambda^2]/\sqrt{\Lambda-1}$.
Although we have not been able to evaluate analytically
the moments $E[\mathcal Y^n]$ with $\mathcal Y = \ln(\mathcal X)$, 
the equations imply that $\mu$ is $O(1/\ln N)$ and
$\sigma$ is $O[1/(\ln N)^2]$.

We can compute $\mu$ using the second method described in 
Sec.~\ref{app:pendulum_a}.
The location of the maximum of Eq.~\ref{eq:F_pi_DW} 
is a solution to a transcendental equation that does not have a closed form 
for arbitrary values of $\Lambda$.
For small positive $\Lambda-1$, however, we find a closed-form solution by 
replacing $\cosh$ in Eq.~\ref{eq:f_X_Lambda} by a constant, chosen such 
that the approximate $P(\mathcal X)$ remains unit normalized.
Thus,
\begin{equation}
    P(\mathcal X) \approx \frac{\Lambda }{\pi (\Lambda-1)} 
    K_0\left( \frac{\Lambda \mathcal X}{2(\Lambda -1)} \right).
    \label{}
\end{equation}
and we find 
\begin{equation}
    \mu \approx \frac{2\pi \sqrt{\Lambda-1}}
    {\ln\left[16N (\Lambda-1)/(\Lambda \varkappa)\right]},
    \label{eq:mu_asym_DW}
\end{equation}
\begin{equation}
    \sigma \approx \frac{\mu^2 }{2 \pi \sqrt{(\Lambda -1 )(1-\varkappa^2)}},
    \label{eq:sig_asym_DW}
\end{equation}
where $\varkappa = 0.595\cdots$ and $\Lambda-1 \ll 1$.

\subsection{Time dynamics of observables}
\label{app:DW_b}

The structure of the phase space of a condensate in double-well potential
is similar to that of the pendulum. Therefore, we can directly apply the 
analysis of time dynamics for a pendulum given in
Sec.~\ref{app:pendulum_b}. In particular, the distribution functions 
$g_{0, R}(\omega_1, \varphi_1)$, as defined in Eq.~\ref{eq:g0R},
are localized and  are given by Eq.~\ref{eq:g0_pend}.
Furthermore, observable $\avg{s_x(t)}$ obeys Eq.~\ref{eq:pndm_O_t_final}.

\section{Spinor gas in single-mode approximation}
\label{app:spinor}

In this  section we obtain results for an antiferromagnetic ($c>0$)
spinor condensate under SMA. 
Its ``single-particle'' Hamiltonian $h_{\rm spin}(\phi, \rho_0)$ is given 
in Eq.~\ref{eq:H_spinor} and $\{\phi, \rho_0 \} =1$. For $-2c<q<0$ the Hamiltonian has a single saddle point 
and a separatrix $S$ dividing the phase space into regions $R=A$ and $B$. In both regions
\cite{zhang_coherent_2005}
\begin{equation}
    \rho_{0,R}(t) = y_2 - (y_2 - y_1) \cn^2(\Omega (t-t_0), k),
    \label{eq:rho0_t}
\end{equation}
where  $\cn(z,k)$ is a Jacobi elliptic function \cite{dlmf} and 
$y_1 \leq y_2 \leq y_3$ are the three real roots of the
cubic equation in $\rho_0$
\begin{equation}
    [\mathcal E -q(1-\rho_0)][(2c\rho_0 + q)(1-\rho_0)-\mathcal E] 
    - (c\rho_0  m)^2 = 0.
    \label{eq:cubic}
\end{equation}
Here, $\mathcal E $ is the ``single-particle'' energy of the trajectory
and $m$ is the unit magnetization.
In terms of these roots,  $\Omega = \sqrt{2|q|c(y_3-y_1)}$ and 
the modulus $k = \sqrt{(y_2 - y_1)/(y_3 - y_1)}$.
The solution is periodic in time with period $T = 2 K(k)/\Omega$ and 
frequency $\omega_1 =2\pi/T= 2\pi\Omega/[2K(k)]$.
 The corresponding  $\phi_R(t)$ is obtained by solving 
 $h_{\rm spin}(\phi_R(t), \rho_{0,R}(t))=\mathcal E$.

On the separatrix $S$  the energy $\mathcal E = 0$ and the
roots of Eq.~\ref{eq:cubic} 
are $y_{1, S} = |q|/(2c)$ and $y_{2,S} = y_{3, S} =1$. Using 
the fact $\cn(x, k) \sim \sech(x)$ as $k \to 1$ and setting $t_0=0$, 
we find the separatrix solution
\begin{equation}
    \rho_{0, S}(t) = 1-(1- y_{1,S})\sech^2(\Omega_S t),
    \label{}
\end{equation}
where $\Omega_S = \sqrt{2|q|c(1-y_{1, S})}$.

\subsection{Distribution function $\mathcal F(\varpi)$}
\label{app:spinor_a}

We now study the distribution $\mathcal F(\varpi)$ for the spinor
condensate by relating the auxiliary frequency $\varpi$ to the
conserved quantities $\cal E$, $m$ and $N$.  As the initial Wigner
distribution $F_0(\psi_j, \psi_j^*)$ is localized near the saddle point
with $\rho_0 = 1$, i.e., $(\psi_{+1}, \psi_0, \psi_{-1}) = (0, \sqrt{N},
0)$, we again define real coordinates $p_{j}$ and $q_j$ via $\psi_j =
\delta_{j 0} \sqrt{N} + p_j + i q_j$.  Then the relevant trajectories
have energy $\mathcal E=0+ \widetilde{\mathcal E}/N + O(N^{-3/2})$ and
unit-magnetization $m=0+  {\widetilde m}/N + O(N^{-3/2})$, both close
to zero.  The quantities $\widetilde{\mathcal E}$ and ${\widetilde m}$
are $O(1)$ and depend on $p_j$ and $q_j$.  We solve for the roots $y_i$
perturbatively with small parameter $1/N$ and find that the modulus $k$
is close to one. Then the auxiliary frequency $\varpi=\omega_1\sim
2\pi\Omega/\ln(16/k'^2)$ in regions $A$ and $B$. We define
\begin{equation}
    \mathcal X \equiv Nk'^2 \sim
     \frac{c}{|q|}\frac{ \sqrt{\mathstrut \widetilde{\mathcal E}^2 + \alpha {\widetilde m}^2}}{(1-y_{1,S})^2},
    \label{}
\end{equation}
which is independent of $N$, and  $\alpha = 2|q|(1- y_{1,S})/c$.
Conversely, $\varpi = 2\pi \Omega/[\ln(16N/\mathcal X)]$.  Unlike
for the previous two systems, we have not been able to
find an analytical expression for the distribution of $\mathcal X$.
Nevertheless, we can apply Eq.~\ref{eq:mu_asym_series} with small
parameter $\lambda^{-1} = \Omega/\ln(16N)$ and find
\begin{equation}
    \mu \sim \frac{2\pi \Om}{\ln(16 N)}.
    \label{eq:mu_spinor}
\end{equation}
Moreover, Eq.~\ref{eq:sig_asym_series} implies that $\sigma = O[1/(\ln N)^2]$; 
hence, $\sigma \ll \mu$ as $N \to \infty$.

\begin{figure}
  \begin{center}
    \includegraphics[scale=0.5]{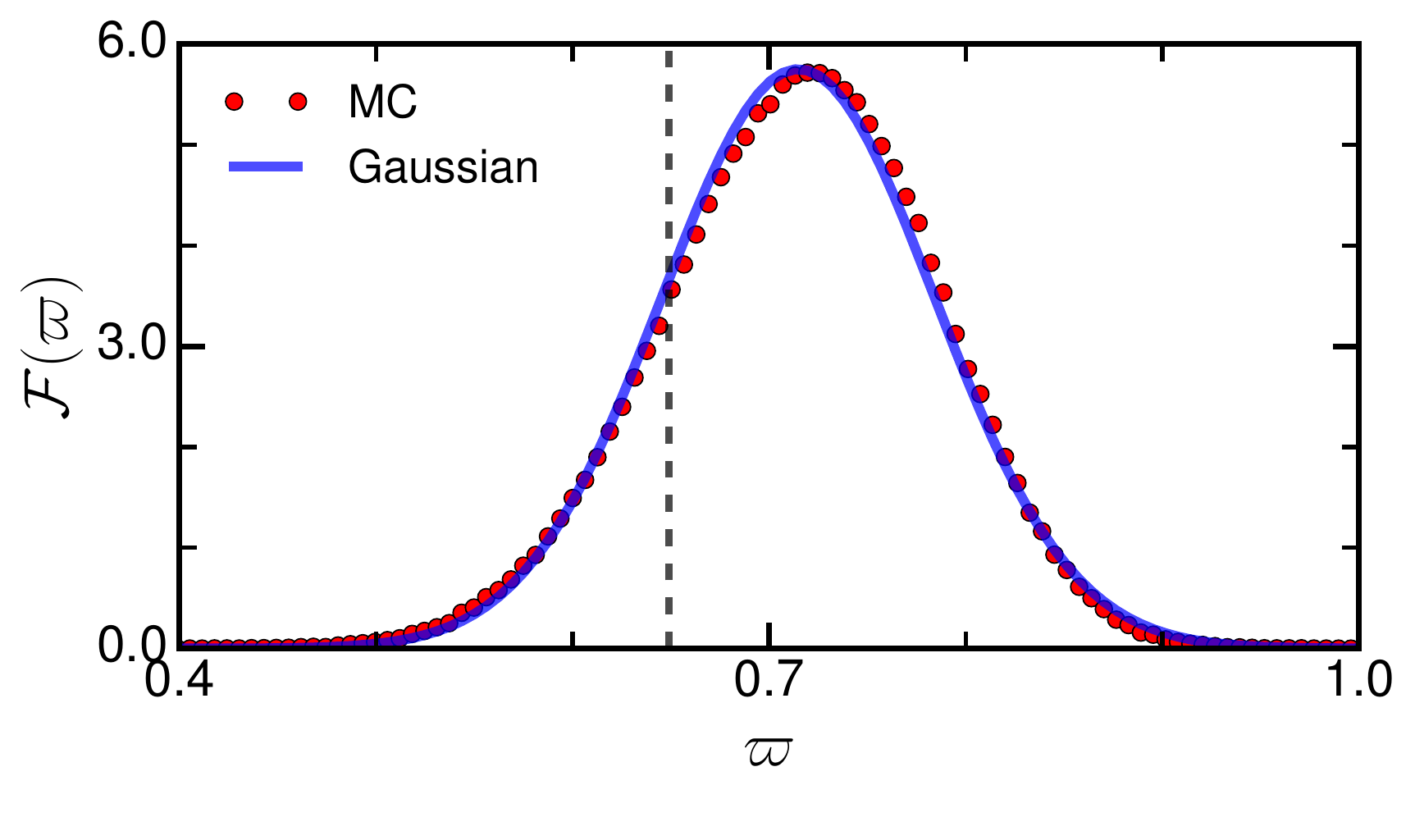}
  \end{center}
  \caption{
Distribution function $\mathcal F(\varpi)$ as a function of the auxiliary frequency $\varpi$
for a spinor condensate with $1000$ atoms and $q/c = -1$.
Red dots represent $\mathcal F(\varpi)$ obtained
by Monte Carlo sampling of the initial Wigner distribution given by
Eq.~\ref{eq:f0_spinor} and
the blue solid line is a Gaussian fit to this data. The mean according 
to Eq.~\ref{eq:mu_spinor} is the dashed vertical line.
  }
  \label{fig:dist_spinor}
\end{figure}

We have numerically evaluated $\mathcal F(\varpi)$ and found that
it is a Gaussian to a good approximation for $-2c < q <0$.  Figure
\ref{fig:dist_spinor} shows $\mathcal F(\varpi)$ for $q/c=-1$ and $N=1000$
and a Gaussian fit to this distribution.  For fig.~\ref{fig:spinor_plot}
we use the mean and width of the numerically obtained 
$\mathcal F(\varpi)$.

\subsection{Time dynamics of observables for $-2c<q<0$}
\label{app:spinor_b}

We now obtain an approximation for $g_{0, R}(\omega_1, \varphi_1)$, 
as defined in Eq.~\ref{eq:g0R}, 
for the spinor system, where $R \in \{A, B\}$.
The initial 
Wigner distribution $F_0(\psi_i, \psi_i^*)$
is localized around the saddle point and, thus,
we expect $g_{0, R}(\omega_1, \varphi_1)$ to be localized around the 
$\varphi_1=0$ (see Fig~\ref{fig:spinorphase}). 
This can be formally justified by writing $\rho_0(t)$
along a trajectory near the separatrix in terms of the angle $\varphi_1$. 
Then, similar to Sec.~\ref{app:pendulum_b_a} 
we can show that the spread in $\varphi_1$ is much smaller than one 
where $g_{0, R}(\omega_1, \varphi_1)$ is significant. Thus, 
\begin{align}
g_{0, R}(\omega_1,  \varphi_1) 
&\approx
2\pi \overline g_{0, R}(\omega_1) \delta(\varphi_1),
  \label{eq:g0_spin}
\end{align}
where $\overline{g}_{0, R}(\omega_1) = \int_0^{2\pi}
d\varphi_1/(2\pi) g_{0,R}(\omega_1, \varphi_1)$ is a marginal distribution.

\subsection{Time dynamics for  $q=0$}
\label{app:spinor_c}

The dynamics of a spinor condensate quenched to $q=0$ is 
qualitatively different  from  that for $q < 0$. 
Instead of a single saddle point, the Hamiltonian has a degenerate line of 
saddle points along $\phi = \pi$.
Along a trajectory close to this line $\rho_0(t)$ is a sinusoid
given by
\begin{equation}
  \rho_0(t) \sim \cos^2[\sqrt{2c \mathcal E}(t + t_0)], 
  \label{eq:rho0_q_0}
\end{equation}
where energy $\mathcal E\equiv h_{\rm spin}(\phi, \rho_0) > 0$ 
and $t_0$ is determined by  the initial condition. This trajectory 
does not spend a significant fraction of its time period near $\rho_0 =1$
that violates one of the assumptions under which 
Eq.~\ref{eq:long_time_avg_final} was derived.

We can, nevertheless, find an analytical expression for $\avg{\rho_0(t)}$
by evaluating the expectation value directly from Eq.~\ref{eq:def_avg}.
The initial Wigner distribution, Eq.~\ref{eq:f0_spinor}, is localized
around $\rho_0=1$, and thus time $t_0 \approx 0$ for the relevant
trajectories. Hence, we only require the distribution function
\begin{equation}
P(\mathcal E)=\int d\psi_i^* d\psi_i\,  F_0(\psi_i,\psi_i^*) 
    \delta(\mathcal E- h_{\rm spin}(\phi, \rho_0)). 
    \label{eq:PofE}
\end{equation} 
Now $\rho_0 =1$ corresponds to the mean-field state
$(\psi_{+1}, \psi_0, \psi_{-1}) = (0, \sqrt{N}, 0)$ and
near $\rho_0 =1$ the Hamiltonian $h_{\rm spin}(\phi, \rho_0) = c[(p_{+1} + p_{-1})^2
+ (q_{+1} - q_{-1})^2]/N + O(N^{-3/2})$, with quadratures 
$p_j$ and $q_j$ defined by $\psi_{+1} = p_{+1} + iq_{+1}$ and $\psi_{-1} = p_{-1} + i q_{-1}$.  
Substituting the Wigner distribution into Eq.~\ref{eq:PofE} and computing the integrals, we
find $P(\mathcal E) \sim N c^{-1}e^{-N \mathcal E/c}$. Finally, averaging
Eq.~\ref{eq:rho0_q_0} over this distribution yields
\begin{equation}
    \avg{\rho_0(t)} \sim 1- \, \alpha t\, F(\alpha t),
\end{equation}
where $\alpha = c \,\sqrt{2/N}$ and $F(x)$ is the Dawson integral \cite{dlmf}.

\bibliography{DI_integrable_few_modes}

\end{document}